\documentclass[twocolumn]{openjournal}

\usepackage{lipsum}

\usepackage{xcolor}
\usepackage{textgreek}
\usepackage[utf8]{inputenc}
\usepackage[english]{babel}

\usepackage{hyperref}
\hypersetup{
    unicode, 
    colorlinks=true,
    linkcolor=linkcolor,
    citecolor=linkcolor,
    filecolor=linkcolor,
    urlcolor=linkcolor,
}
\usepackage{color,colortbl}
\definecolor{linkcolor}{rgb}{0.0,0.3,0.5}
\usepackage{tensind}
\tensordelimiter{?}
\DeclareGraphicsExtensions{.bmp,.png,.jpg,.pdf}
\usepackage{verbatim}
\usepackage[normalem]{ulem}
\usepackage{orcidlink}
\usepackage{soul}
\usepackage{graphicx}
\usepackage{booktabs}
\usepackage{amsmath,amssymb}
\usepackage{gensymb}
\usepackage{soul}
\usepackage{multirow} 

\graphicspath{ {./figs/} }

\begin{document}
\title{IRMaGiC: Extending Luminous Red Galaxy Selection into the Infrared with Joint Rubin Observatory's Large Survey of Space Time and Roman's High Latitude Imaging Survey}

\author{Zhiyuan Guo\orcidlink{0000-0001-9557-9171}}
\affiliation{Department of Physics, Duke University, Durham, NC 27708, USA}

\author{C. W. Walter\orcidlink{0000-0003-2035-2380}}
\affiliation{Department of Physics, Duke University, Durham, NC 27708, USA}

\author{Eli S. Rykoff\orcidlink{0000-0001-9376-3135}}
\affiliation{SLAC National Accelerator Laboratory, Menlo Park, CA 94025, USA}
\affiliation{Kavli Institute for Particle Astrophysics and Cosmology, 
P.O. Box 2450, Stanford University, Stanford, CA 94305, USA}

\author{The LSST Dark Energy Science Collaboration}

\begin{abstract}

We introduce IRMaGiC, an algorithm built on RedMaGiC designed to enhance the selection of Luminous Red Galaxies (LRGs) across the redshift range $1 \leq z \leq 2$. We show that this method extends the capabilities of the RedMaGiC algorithm by applying it to simulated photometric data from the Vera C. Rubin Observatory’s Legacy Survey of Space and Time (LSST) and the Nancy Grace Roman Space Telescope’s High Latitude Imaging Survey (HLIS). By integrating infrared band coverage from Roman with LSST's optical bands, IRMaGiC enables red-sequence calibration at higher redshifts. We demonstrate that IRMaGiC reduces scatter and bias in photometric redshift estimates for LRGs at higher redshift, providing more accurate redshift assessments compared to existing methods. Our findings suggest that incorporating infrared data can considerably improve the selection and redshift estimation of LRGs at higher redshift, offering substantial benefits for future cosmological surveys.

\end{abstract}

\maketitle

\section{Introduction}

Luminous red galaxies (LRGs) represent a fundamental component of the cosmic landscape, playing a pivotal role in the study of cosmological evolution and structure formation. These galaxies, characterized by their lack of significant star formation and their red colors due to older stellar populations, serve as important tracers of the mass assembly history of the universe~\citep{Choi14,Khullar22,Marsan22,Zhuang23,Beverage24,Slob24}. Understanding the distribution and properties of LRGs across cosmic time provides insights into the mechanisms that drive galaxy evolution and the influence of environment on these processes. In cosmology, red, quiescent galaxies are often utilized as proxies to identify high-density environments like galaxy clusters and groups. The presence of the 4000\AA~break in their spectra makes them ideal candidates for accurate redshift estimation. Due to their brightness and distinct spectral features, these galaxies are easily detectable over a wide range of redshifts, making them ideal probes for studying the large-scale structure of the universe, with such probes as Baryon Acoustic Oscillation (BAO)~\citep{Percival07,Eifler21,Rosell22,DESI3} and Galaxy Clustering (GC)~\citep{Padmanabhan07,Pandey22,White22,Sailer24,Yuan24}.

There are various methods for selecting luminous red galaxies (LRGs) from a given survey. Surveys such as the Sloan Digital Sky Survey (SDSS)~\citep{Eisenstein01} and the Dark Energy Spectroscopic Instrument (DESI)~\citep{Zhou23a,Zhou23b} use direct color cuts with different color combinations to identify LRG candidates across different redshifts. An alternative method is the red-sequence technique, which leverages the characteristic that predominantly red, luminous, and passively evolving red-sequence galaxies—known for their aged, metal-rich stellar populations—form a distinct ridge line~\citep{Visvanathan77} at a specific redshift. This indicates that their colors vary minimally with magnitude in the color-magnitude diagram. By using a sample of red galaxies with known spectroscopic redshifts, one can calibrate and model the color-redshift evolution of the red-sequence, known as the red-sequence template. This technique was initially developed for cluster-finding algorithms through the identification of LRGs using multi-band photometry. For example, the RedMaPPer~\citep{Rykoff14,Korytov19} algorithm uses spectroscopic samples from SDSS to iteratively fit a red-sequence template to identify clusters in SDSS and the Dark Energy Survey (DES)~\citep{DES16}. Similarly, the CAMIRA~\citep{Oguri14,Oguri17} algorithm employs stellar population synthesis models, also calibrated using SDSS data, to predict the colors of red-sequence galaxies and identify clusters in the Subaru Hyper Suprime-Cam survey (HSC)~\citep{Aihara18}. 

Further advancements have built upon the calibrated red-sequence template to create an effective red galaxy selection algorithm based solely on photometric data, enabling the selection of LRGs in galaxy samples while simultaneously providing accurate photometric redshift estimates. For example, the RedMaGiC algorithm~\citep{Rozo16} builds upon the calibrated red-sequence template from RedMaPPer to select and generate LRG catalogs for SDSS and DES. As mentioned earlier, in this algorithm the red-sequence is initially calibrated using cluster-finding algorithms because red galaxies typically reside at the centers of clusters, and neighboring galaxies within the same cluster are likely to share the same redshift as the central red galaxy. This method thus leverages the red central galaxies to extend the red-sequence calibration to fainter magnitudes, thereby increasing the number of data points available for the calibration process. However, with enough spectroscopic galaxy samples, the red-sequence template can be also calibrated independently without relying on galaxy clusters. For example, \citet{Vakili19} calibrated a red-sequence template for the Kilo-Degree Survey (KiDS)~\citep{deJong13} Data Release 3~\citep{deJong17} using ugri photometry and spectroscopic galaxies from the overlapping regions of The Galaxy and Mass Assembly (GAMA) survey~\citep{Driver11} and SDSS. In follow-up work, \citet{Vakili23} extended this calibration using KiDS DR4, incorporating The VISTA Kilo-degree Infrared Galaxy (VIKING) Survey~\citep{Edge2013} ZYJHKs near-infrared bands. This calibrated template was then used to generate RedMaGiC-like galaxy samples from KiDS with similar photometric redshift performance. However, most previous surveys have been limited to probing LRGs up to $z \lesssim 1$ due to the constraints of optical band coverage, as the $4000,\mathrm{\AA}$ break shifts into the near-infrared at higher redshifts. More recently, \citet{Oguri26} extended the CAMIRA LRG sample to $z \sim 1.25$. However, this extension remains limited by the lack of near-infrared coverage, as the reddest band available in HSC-SSP is the $y$ band.

\begin{figure}[t]
    \centering
    \epsscale{1.2}
    \plotone{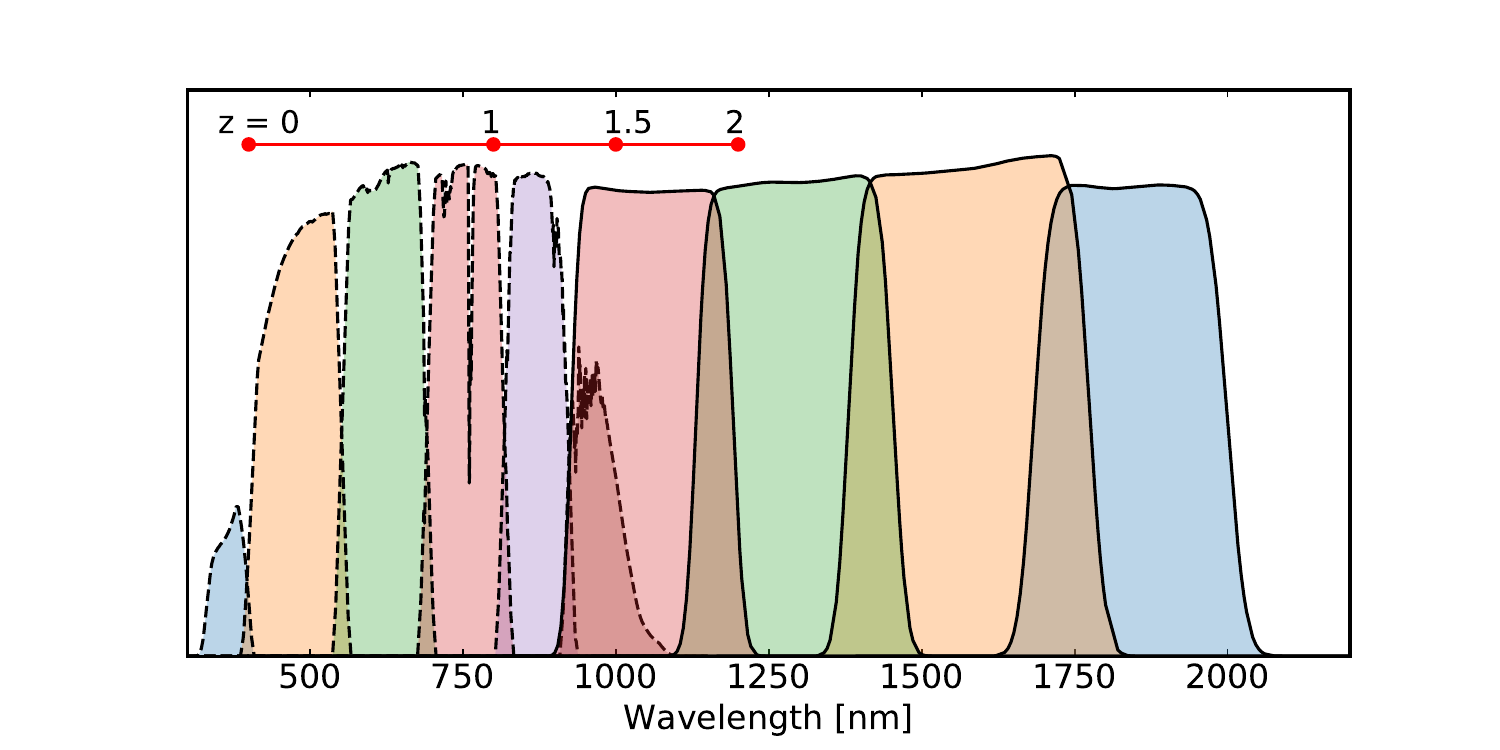}
    \caption{Throughput curves for the six LSST filters and four Roman filters. From left to right, the LSST $u,\ g,\ r,\ i,\ z,\ y$ filters are represented by black dashed lines, while the Roman $F106,\ F129,\ F158,\ F184$ filters are shown with black solid lines. The red solid horizontal line marks the wavelength of the $4000\mathrm{\AA}$ break as the galaxy spectra get redshifted. This illustration is intended to demonstrate the overall shape and wavelength coverage of the filters and does not represent the actual throughput values.}
    \label{fig:LSST+Roman_bands}
\end{figure}

In this paper, we introduce \textbf{IRMaGiC}, the InfraRed extension of RedMaGiC, and apply it to simulated datasets from two Stage IV dark energy surveys as categorized by the Dark Energy Task Force~\citep{Albrecht06}: the Vera C. Rubin Observatory's Legacy Survey of Space and Time~\citep{Ivezic19} (hereafter referred to as LSST) and the Nancy Grace Roman Space Telescope's High Latitude  Wide Area Survey ~\citep{Spergel15, Akeson19,Roman25} (hereafter referred to as Roman HLWAS). Our goal is to adapt a RedMaGiC-like algorithm for selecting Luminous Red Galaxy (LRG) samples within the redshift range of $1 \leq z \leq 2$. This is made feasible because Roman HLWAS includes infrared wavelength coverage, complementing the optical bands of LSST as shown in Figure~\ref{fig:LSST+Roman_bands}. Thus, it becomes possible to extend the red-sequence template to higher redshifts. Our approach is similar to that of~\citet{Vakili19} and~\citet{Vakili23}, as we rely on a sufficient supply of spectroscopically confirmed red galaxies from the spectroscopy component of the Roman HLWAS, the High Latitude Spectroscopic Survey (hereafter HLSS), for red-sequence template calibration.

This paper is organized as follows: In Section~\ref{sec:data}, we describe the simulated datasets used in this study for LSST, Roman High Latitude Imaging Survey (HLIS), and HLSS. Section~\ref{sec:method} provides an overview of the IRMaGiC algorithm, including red-sequence calibration (Section~\ref{sec: rs-caliberation}), the implementation of the Roman HLSS red galaxy efficiency curve (Section~\ref{sec: Grism efficiency}), and the candidate selection process (Section~\ref{sec: red galaxy selection}). We also introduce a novel method of identifying high-redshift ($z > 1$) red galaxy candidates for red-sequence calibration in Section~\ref{sec: seed galaxy selection}. In Section~\ref{sec:results}, we evaluate the photometric redshift performance of our LRG samples and compare with those obtained from existing photometric redshift estimation codes. We also compare our LRG samples with existing LRG samples at lower redshifts, specifically the cosmoDC2/DC2 RedMaGiC catalogs\footnote{\url{https://github.com/LSSTDESC/gcr-catalogs}} in Section~\ref{sec:results}. Additionally, we investigate how different assumptions about the area of overlap regions between LSST and Roman HLIS affect the maximum achievable redshift range with our red-sequence template.  Finally, we summarize our findings in Section~\ref{sec:summary}.

\section{Data} \label{sec:data}

\subsection{LSST DC2}

The LSST is a ground-based photometric survey designed to observe approximately $20,000\deg^2$ of the sky during its 10-year mission across $ugrizy$ bands~\citep{LSST09}. The survey consists of two main components: the Wide-Fast-Deep (WFD) survey, covering $18,000\deg^2$, and the Deep Drilling Fields (DDFs). This study focuses on the WFD, which is particularly suited for dark energy probes such as weak lensing, galaxy clustering, and Type Ia supernovae~\citep{LSST18}. Given the unprecedented scale of data production by LSST, there is a crucial need to develop and verify software pipelines for analyzing the released data. To address this, the LSST Dark Energy Science Collaboration (DESC) has initiated a series of data challenges aimed at developing an end-to-end simulation pipeline to generate LSST-like data products~\citep{DESC12}. The LSST simulation used in this study is based on the second data challenge (DC2)~\citep{LSSTDC221}. The simulated DC2 sky survey replicates LSST observations over six optical bands $ugrizy$ within a WFD area of approximately $300\deg^2$ and a DDF area of $1\deg^2$, reflecting the expected depth over five years of the planned 10-year survey.

The DC2 simulation starts with the large-volume N-body cosmological simulation, Outer Rim~\citep{Heitmann19} to generate a realistic extragalactic catalog, cosmodc2~\citep{Korytov19}, which includes galaxy properties such as shapes, redshifts, and magnitudes. Instance catalogs for specific telescope pointings are then used to create synthetic images using the DESC image simulation package IMSIM\footnote{\url{https://github.com/LSSTDESC/imSim}}, which simulates telescope and atmospheric effects. These images are subsequently processed through the LSST Science Pipelines to produce data products analogous to those expected from the actual LSST survey. For additional information, we direct the readers to~\citet{LSSTDC221,LSSTDC2DR}.

In this study, we use the static object catalog, \verb|dc2_object_run2.2i_dr6a_with_photoz|, which processes data up to year 5 (DR5) of LSST and provides necessary information such as measured photometry, detected position (RA, DEC) and object types. The catalog is accessed via the \verb|GCRcatalog| interface\footnote{\url{https://github.com/LSSTDESC/gcr-catalogs}}. We use the measured composite model (cModel) photometry as recorded in the catalog for galaxy magnitudes and corresponding errors. The galaxy/star separation is done by filtering the catalog via the \verb|extendness| parameter ranging from 0 to 1, which indicates whether the light from an object is concentrated (point-like) or spread out (extended). In this study, we select objects with \verb|extendness|$\ > 0.5$ as galaxy candidates from LSST DC2.

\subsection{Roman simulation} \label{sec: Roman data description}
The Roman High Latitude Survey consists of two components: an imaging component, the High Latitude Imaging Survey (HLIS), and a spectroscopy component, the High Latitude Spectroscopy Survey (HLSS). Based on the current Roman reference survey~\citep{Troxel23,Wang22}, the HLIS and HLSS will observe a 2000 deg$^2$ region of the sky, fully within the LSST footprint, over a five-year period in four near-infrared bands: $F106,\ F129,\ F158,\ F184$, enabling joint analyses between the two surveys. In this paper, we denote these four Roman bands as $Y,\
J,\ H,\ F$.

The Roman galaxy sample used in this study is derived from the joint Roman HLIS-LSST image simulation presented in~\citet{Troxel23}. This simulation is based on the updated Roman HLIS image simulation pipeline, which includes a realistic Point Spread Function (PSF), sensor physical effects, and chromatic rendering~\citep{Troxel23}. The simulation focuses on a 20 deg² region of the LSST-DC2 universe, spanning $51^\circ < \mathrm{RA} < 56^\circ$ and $-42^\circ < \mathrm{DEC} < -38^\circ$, for the five-year depth of the HLIS. This joint simulation provides an excellent testbed for joint survey analysis. It uses the same basic truth input as the DC2 simulation, with adjustments specifically tailored for the need of Roman HLIS. For further details, the readers are referred to~\citet{Troxel23}.

Object detection and measurements are performed on coadded images in each HLIS band using SourceExtractor~\citep{SExtractor}. In this study, we use the provided \verb|mag_auto| values for galaxy magnitudes and their associated errors in each Roman band.  For the Roman simulated data, we rely on the truth information to select galaxies. We cross-reference the detection catalog generated by SourceExtractor with the input truth galaxy catalog by matching objects based on their celestial coordinates, which is further refined by selecting those with the closest photometry. It is important to note that the Roman science pipeline has not yet been finalized, so there may be discrepancies between the results presented here and those from future analyses, and we defer the star/galaxy separation for the Roman HLIS to a future study.

\subsection{Cross matching}
A combined galaxy sample from both surveys is essential to calibrate a red-sequence model for the targeted redshift range ($1 < z < 2$). Objects that show significant detections in the optical bands may not necessarily have strong detections in the near-infrared bands, and vice versa. Therefore, crossmatching the galaxy catalogs from the two surveys is crucial to ensure that the galaxies in the combined catalog have significant detections in both. This crossmatching is performed based on positional (RA, DEC) alignment, retaining matches within 0.2 arcseconds and requiring significant detections ($>5\sigma$) in the Rubin LSST i, z, and Roman Y, J, H, and F bands. This is accomplished by setting the following parameter thresholds on the DC2 catalog: \verb|snr_i_cModel| $\geq 5$ and \verb|snr_z_cModel| $\geq 5$, and by ensuring \verb|flux_auto|/\verb|flux_auto_err| $\geq 5$ in the Roman catalog. In this study, we focus primarily on relatively bright galaxies. To verify the positional matching results, we cross-checked with information from the truth catalogs of the DC2 and Roman simulations. We find that for galaxies with $m_{\mathrm{H}} > 23$, $99\%$ have truth matches based on galaxy id, confirming that the cross-matched sources from both catalogs represent the same galaxies.

\subsection{Roman Grism spectroscopy}
As previously noted, a sample of spectroscopically confirmed red galaxies in the NIR/IR spectrum is essential for precise calibration of the red-sequence template at higher redshift. In this study, we assume that the spectroscopic galaxy sample will be sourced from  Roman HLSS. The Roman grism enables spectroscopy at a resolution of $R = 460$ for wavelengths in the range $\lambda = 1 - 1.93 \mu m$. In \citet{Guo24}, we conducted a detailed analysis of the Roman grism's capability to measure spectroscopic redshifts (spec-z) for red, quiescent galaxies via simulation. Then we predicted the expected redshift recovery rate for these galaxies observed in the Roman HLSS. Further details are discussed in Section~\ref{sec: Grism efficiency}.

\section{Method} \label{sec:method}

\subsection{Algorithm overview}

Red-sequence galaxies display a consistent color-redshift relationship, aligning along a defined ridge in the color-magnitude diagram within a certain intrinsic scatter at any given redshift. This feature can be characterized empirically by the following red-sequence template (thereafter RS-template), which is modeled through a set of the parameters smoothly evolving as a function of redshift:
\begin{equation} \label{eq:rs-model}
    \langle \textbf{c}|z,m\rangle = \textbf{a}(z) + \textbf{s}(z)[m-{m_{\mathrm{ref}}}(z)]
\end{equation}
where $z$ is the redshift, \textbf{c} is the color vector {$i - z, z - Y, Y - J, J - H, H - F$} of a galaxy, \textbf{a}($z$) and \textbf{s}($z$) are the redshift-dependent intercept (mean color) and slope of the ridge-line. $m$ is the reference magnitude in the reference band. According to \citet{Rozo16}, the reference band should lie redwards of the 4000$\mathrm{\AA}$ break. In this study, we choose $F$ as the reference band.  $m_{\mathrm{ref}}(z)$ is defined as the pivot point of the color-magnitude relation and it can be chosen arbitrarily. In this study, we follow \citet{Rozo16} and set it as the median magnitude of the seed galaxies in the reference band. Additionally, a color covariance matrix, $\mathbf{C}_{\textbf{tot}}$ is defined for every galaxy.
\begin{equation} \label{eq:cov}
    \mathbf{C}_{\textbf{tot}} = \mathbf{C}_{\textbf{obs}} + \mathbf{C}_{\textbf{int}}(z)
\end{equation}
It is composed of two component $\mathbf{C}_{\textbf{obs}}$, the observed color covariance, for any color pair with $C_i = m_a - m_b$ and $C_j = m_q - m_p$,
\begin{equation}
\mathbf{C}_{\textbf{obs}}(z) = 
    \begin{pmatrix}
    \sigma^2_a + \sigma^2_b & \gamma  \\
    \gamma & \sigma^2_q + \sigma^2_p
    \end{pmatrix}     
\end{equation}
with
\begin{equation}
    \gamma = \begin{cases} -\sigma^2_q & \text{if } b = q \\ 0 & \text{if } b \neq q \end{cases}
\end{equation}
and $\mathbf{C}_{\textbf{int}}(z)$, the intrinsic red-sequence color covariance. The calibration of the red-sequence parameters $\mathbf{a}(z)$, $\mathbf{s}(z)$ and $\mathbf{C_{\mathrm{int}}(z)}$ is discussed in detail in Section~\ref{sec: rs-caliberation}.

The goal is to estimate the photometric redshift of a galaxy with a certain magnitude $m$ and color vector $\mathbf{c}$ with a calibrated  RS-template. The redshift probability of a red galaxy with H band magnitude $m$, and color vector $\mathrm{c}$ can be calculated as:
\begin{align*}
    P(z|\mathbf{c},m) &= \frac{P(\mathbf{c},m,z)}{P(\mathbf{c},m)} \\
                      &\propto P(\mathbf{c}|m,z)P(m|z)P(z)
\end{align*}
We are interested in calculating the likelihood
\begin{equation}
    \mathcal{L} = P(\mathbf{c}|m,z)P(m|z)P(z)
\end{equation}
The first term $P(\mathbf{c}|m,z)$ is defined as:
\begin{equation}
    P(\mathbf{c}|m,z)\propto \exp{\left[-\frac{1}{2}\chi_{\mathrm{red}}^2(z)\right]}
\end{equation}
where
\begin{equation} \label{eq: chi2_red}
    \chi_\mathrm{red}^2(z) = (\mathbf{c}-\langle \textbf{c}|z,m\rangle)^T\mathbf{C^{-1}_{\mathrm{tot}}}(\mathbf{c}-\langle \textbf{c}|z,m\rangle)
\end{equation}
and $\mathbf{C_{tot}}$ is the galaxy color covariance defined above in Equation~\ref{eq:cov}.
The second term $P(m|z)$ serves as a redshift-dependent luminosity filter and is modeled assuming the galaxies follow a Schechter luminosity function:
\begin{align*}\label{eq:lum_dist}
    P(m|z) &\propto  \\
    &10^{-0.4(m-m_{\ast}(z))(\alpha+1)}\exp{\left[-10^{-0.4(m_i-m_{\ast}(z))}\right]}
\end{align*}
Following~\citet{Rozo16}, the parameter $\alpha$ is set to $1$. $m_{\ast}(z)$ is the characteristic magnitude in the reference band. And it is computed using  the Bruzual\ \& Charlot stellar population synthesis code~\citep{BC03} (hereafter referred to as BC03) implemented in the \verb|EzGal| Python package\footnote{\url{https://github.com/cmancone/easyGalaxy}} assuming a solar metallicity, a Salpeter initial mass function~\citep{Salpeter55}, and a single star formation burst at $z = 3$ following~\citet{Rozo16}. The normalization condition is computed from the BC03 model using the Roman F bandpass.
Finally, the last term, $P(z)$, the redshift prior, is defined as the derivative of the comoving volume with respect to redshift. 
\begin{equation}
    P(z)\propto\frac{dV}{dz} = (1+z)^2D^2_A(z)cH^{-1}(z)
\end{equation}
where $c$ is the speed of light, $H(z)$, $D_A(z)$ are the Hubble parameter and the angular diameter distance as a function of redshift.
It is important to note that this prior is implemented solely for the purpose of redshift fitting in the likelihood function. We do not apply a constant comoving density constraint to the final RedMaGiC sample, as will be discussed in Section~\ref{sec: red galaxy selection}.
Thus, the final expression for the redshift likelihood function is expressed as:
\begin{align} \label{eq:likelihood}
\begin{split}
        \ln\mathcal{L}(z) = &-\frac{1}{2}\chi_\mathrm{red}^2(z) - \frac{1}{2}\ln\ {\mathrm{det}(\mathbf{C}_{\mathrm{tot}}(z))} \\
                        &+ \ln P(m|z) + \ln\left|\frac{dV}{dz}\right|
\end{split}
\end{align}
The photometric redshift $\widetilde{z_{\mathrm{red}}}$ is computed as the integral of the likelihood function over the redshift range:
\begin{equation}
    \langle\widetilde{z_{\mathrm{red}} }\rangle = \frac{\int dz\ \mathcal{L}(z)z}{\int dz\ \mathcal{L}(z)}
\end{equation}
The tilde is used to indicate that this value is an initial estimate of IRMaGiC photo-z, which is further refined using an afterburner (see Section~\ref{sec: afterburner}).
The corresponding photometric redshift error $\sigma_z$ is computed as:
\begin{equation}
    \sigma^2_z = \langle z^2\rangle - \langle z\rangle^2
\end{equation}
Additionally, for each galaxy, we assigned a $\chi^2_{\mathrm{red}}$ value, corresponding to the chi-square value at the redshift that maximizes the log-likelihood function in Equation~\ref{eq:likelihood}, and a luminosity ratio,
\begin{equation} \label{eq:lum-ratio}
    l(m,\widetilde{z_{\mathrm{red}}}) = \frac{L}{L_\ast} = 10^{-0.4(m-m_\ast(\widetilde{z_\mathrm{red}}))}
\end{equation}
where $m_*$ and $L_*$ are the characteristic magnitude and luminosity as discussed above. 

$\chi^2_{\mathrm{red}}$ indicates how well a galaxy's photometry is fitted by the calibrated red-sequence template (See Section~\ref{sec: rs-caliberation}), while luminosity ratio is used to determine whether a galaxy satisfies the luminosity threshold (See Section~\ref{sec: red galaxy selection}).  The two quantities are estimated for all detected galaxies within the simulation region and are later used for LRG candidate selection (See Section~\ref{sec: red galaxy selection}).

\subsection{Seed galaxy selection} \label{sec: seed galaxy selection}

\begin{figure*}
    \centering
    \plotone{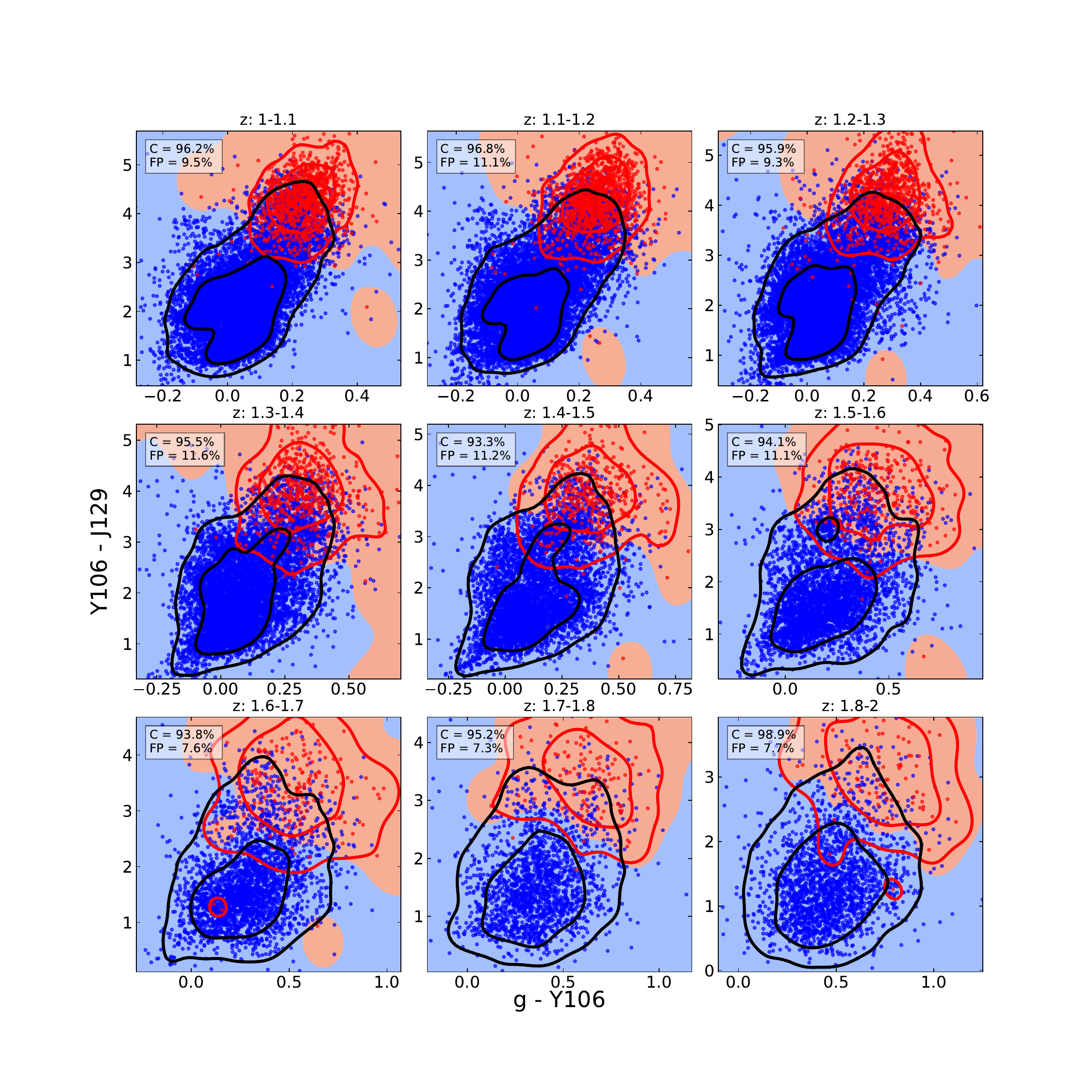}
    \caption{The observed ($Y - J$) versus ($g - Y$) colors. Data are shown at different redshift bins, from $z = 1$ to $z = 2$. Galaxies are color coded depending on their sSFR and Red-sequence flag based on truth information. The red dots are quiescent galaxies and blue dots are star-forming galaxies. The black and red solid lines show the 68 and 95 percent contours of the number density of star-forming and quiescent, red galaxies respectively. The red and blue contours show the classification boundary from SVM. On the top left-hand side of each panel, we report the completeness (C) and False-positive (FP) fraction of the red, quiescent galaxy selection.}
    \label{fig:blue-red-gal-spe}
\end{figure*}

\begin{figure*}
    \centering
    \epsscale{1.15}
    \plotone{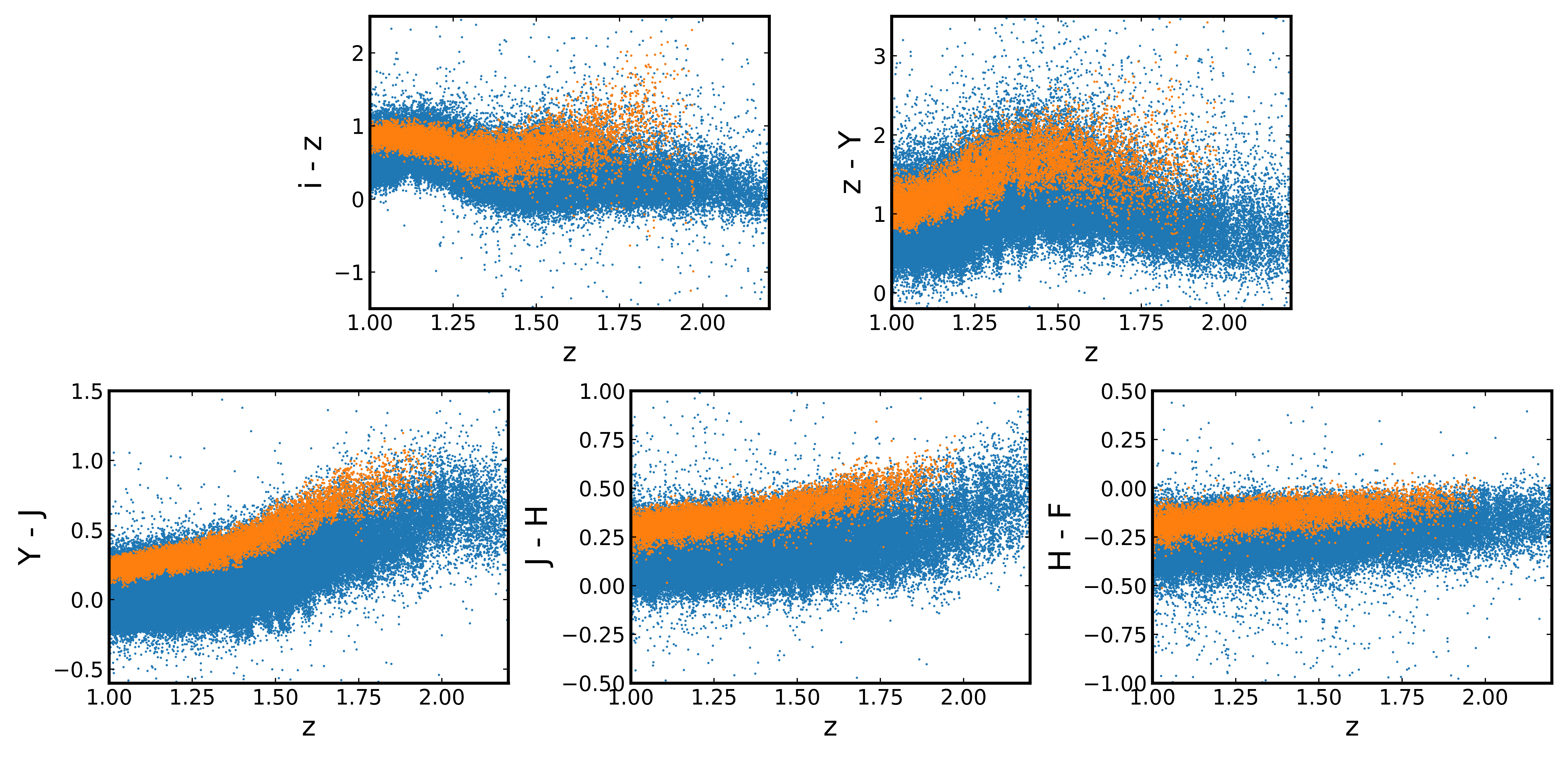}
    \caption{Color evolution versus true redshift for all galaxies with $m_\mathrm{H} \leq 22\ \mathrm{mag}$ cross-matched from the 20 deg$^2$ region of the Roman simulation and LSST DC2. The blue points represent all galaxies included in this study, while the orange points indicate the seed galaxies selected for calibrating the red-sequence template. The five panels illustrate the redshift dependence of the $i - z$, $z - Y$, $Y - J$, $J - H$, and $H - F$ colors. Here, the $i$ and $z$ band magnitudes are sourced from LSST-DC2 measurements, whereas the $Y$, $J$, $H$, and $F$ magnitudes are derived from the Roman simulation.}
    \label{fig:c_vs_z_plot_seed}
\end{figure*}

A set of LRGs with accurate spectroscopic redshift (spec-z) is essential for calibrating the red-sequence. This set of galaxies is known as the seed galaxy sample. Previous studies have selected LRGs using color cuts~\citep{Eisenstein01, Oguri18b,Crocce19, Zhou23} or by applying a Gaussian mixture model~\citep{Rykoff16, Vakili19} within narrow redshift bins ($\Delta z \approx 0.02$), leveraging the color bimodality of the two populations to differentiate between red and blue galaxies. However, these methods are primarily effective for galaxies at low redshifts ($z < 1$) and are not suitable for this study for two main reasons. First, the distinction between blue and red galaxies diminishes at higher redshifts, rendering a direct cut in color space ineffective. Second, although the Roman HLSS will provide spectroscopic redshifts for galaxies, the efficiency curve of LRGs significantly diminishes for fainter cases. This efficiency decline implies that when galaxies with spectroscopic redshifts are segmented into narrow bins, the distribution will be predominately composed of blue populations. Consequently, it becomes challenging to assemble a sufficiently large sample of red galaxies within these narrow bins to enable a meaningful Gaussian mixture model that can effectively characterize their distribution.

The Euclid Space Telescope~\citep{Euclid25} (hereafter Euclid)\footnote{\url{http://sci.esa.int/euclid/}} is another deep-space imaging survey operating at optical and near-infrared wavelengths, featuring broad-band filters, $Y_E, J_E, H_E$, similar to Roman. We use the subscript 'E' to differentiate these from the Roman bands. For Euclid, \citet{Bisigello20} have developed a color-color cut criteria to separate quiescent galaxies from star-forming galaxies up to $z = 2.5$ using color combinations such as, $VIS - Y_E$, $J_E - H_E$ and $u - VIS$, $VIS - J_E$, where $VIS$ refers to the Euclid Visible Instrument, and 'u' represents the u-band from the Canada–France Imaging Survey (CFIS)~\citep{Gwyn25}, which serves as the complementary ground-based survey for Euclid in the north. In this study, we adopt a similar strategy, combining the near-infrared filters from Roman with the optical bands from LSST to select seed red-sequence galaxies. We leverage truth information from simulated data, such as star formation rates and red-sequence flags, to label red and blue galaxy populations. These labels, along with the corresponding photometric measurements, are used to train decision boundaries, which can then be applied to future real data where ground-truth information is not available.

The first step is to distinguish between quiescent and star-forming galaxies. Given that the efficiency curve of the Roman grism for red galaxies approaches zero at $m_{\mathrm{H}} = 22$~\citep{Guo24}, we apply a magnitude cut on the joint galaxy sample, focusing only on galaxies with $m_{\mathrm{H}} \leq 22$. This magnitude cut discards fainter galaxies, facilitating a clearer identification of separations between the two populations. In literature, many works~\citep{Paspaliaris23,Florez20,Stefanon13,Fontanot09} separated quiescent galaxies from star-forming populations based on a fixed threshold in specific star formation rate (sSFR), calculated as follows:
\begin{equation}
    \mathrm{sSFR} = \frac{\mathrm{SFR}}{M_{\mathrm{star}}}
\end{equation}
where SFR is the stellar formation rate and $M_{\mathrm{star}}$ is the stellar mass. Following those studies, we adopt an sSFR threshold of quiescent galaxies:
\begin{equation}
    \log_{10}({\mathrm{sSFR}}/{\mathrm{yr^{-1}}}) \leq -11
\end{equation}
while star-forming galaxies have
\begin{equation}
    \log_{10}({\mathrm{sSFR}}/{\mathrm{yr^{-1}}}) > -11
\end{equation}
Here we use the quantities \verb|totalStarFormationRate| and \verb|stellar_mass| provided in the DC2 catalog to calculate an sSFR for every galaxy. Moreover, the DC2 truth catalog provides a 'red-sequence' flag for each galaxy. We further refine our sample of quiescent galaxies by selecting those with the \verb|is_on_red_sequence| flag set to True, ensuring a purer sample of red, passively evolved galaxies.
We then bin quiescent and star-forming galaxies into redshift bins to explore the effectiveness of various color combinations available from LSST and Roman in distinguishing quiescent galaxies from the star-forming population in each redshift bin. It is important to note that this binning assumes the availability of only photometric redshifts (photo-z) from LSST for all galaxies. This assumption is made because, in reality, spectroscopic redshifts (spec-z) will not be available for all observed red galaxies, resulting in an insufficient number of red galaxy samples to establish a clear separation between the two populations. Therefore, we propose a strategy where we initially utilize photometric data from both surveys and use photo-z from LSST to galaxies in wide redshift bins $\Delta_z = 0.1$ to select red galaxy candidates, then subsequently using candidates with spec-z from the Roman HLSS as the seed galaxies to calibrate the red-sequence template. According to the LSST official documents\footnote{\url{https://dmtn-049.lsst.io/DMTN-049.pdf}}, the minimum performance targets for photo-z should result in a standard
deviation of $z_{\mathrm{true}}-z_{\mathrm{photo}}$ of $\sigma_z \leq 0.05(1+z_\mathrm{photo})$. We thus compute mock photo-z based on the true redshift using:
\begin{equation}
    z_{\mathrm{photo}} = z_{\mathrm{true}} + \epsilon
\end{equation}
where $\epsilon \sim N(0,\sigma^2)$ and $\sigma = 0.05(1+z_\mathrm{true})$. 

We find that the color combination of ($g-Y$) and ($Y-J$) offers the most robust separation between the two populations. The nine panels in Figure~\ref{fig:blue-red-gal-spe} show the ($g-Y$) versus ($Y-J$) distribution of star-forming and quiescent galaxies in each redshift bin. The black and red solid lines in each panel show the 68 and 95 percent contours of the number density of the two populations. It is important to note that due to the scarcity of red galaxies beyond $z=1.8$, we use a wider bin of $\Delta_z = 0.2$ for the final bin, compared to $\Delta_z = 0.1$ used for the earlier bins. The well-separated $68\%$ contours indicate the presence of the two galaxy populations in this color–color space. However, the $95\%$ contours reveal some overlaps, suggesting that quiescent galaxies remain partially contaminated by the star-forming population. 

Comparing to the approach of \citet{Bisigello20}, who implement a linear boundary in the color-color space to separate the two populations, we have adopted a more robust nonlinear approach. This is achieved using Support Vector Machine (SVM) with a radial basis function (\verb|rbf|) kernel to estimate a nonlinear decision boundary in the 2D color-color space. The SVM model includes two hyper-parameters: \textbf{C}, the regularization parameter, and \textbf{gamma}, the kernel coefficient. We optimized these parameters by exploring the best combination on a parameter grid of $C \in [0.01, 0.1, 1, 10]$ and $\gamma \in [0.1, 1]$ with a step size of 0.1. Optimization was conducted using the GridSearchCV function in \verb|scikit-learn|~\citep{scikit-learn}, employing a custom scoring function defined as $\mathrm{C}(1-\mathrm{FP})$ following \citet{Bisigello20}. Here, C represents the completeness, defined as the fraction of true quiescent and red galaxies correctly classified within the decision boundary, while FP denotes the false-positive fraction, representing the proportion of star-forming, blue galaxies incorrectly classified within the boundary. We constrained the range of $\gamma$ to smaller values in purpose to prevent over-fitting. The final decision boundaries are delineated by blue and red regions in Figure~\ref{fig:blue-red-gal-spe}. Additionally, the completeness and false-positive fraction, given the boundary, are displayed within each panel. It is evident that this approach allows us to achieve an high completeness of approximately $95\%$ with relatively low FP fraction ($\approx10\%$). 

However, it is important to note that the effectiveness of this method may vary across different galaxy samples, as we have implemented a stringent magnitude cut in our analysis based on the anticipated performance of the Roman grism for measuring spectroscopic redshifts of red galaxies. Additionally, variations in the photometry measurements could influence the decision boundary. Therefore, it is necessary to retrain the SVM boundaries before applying them to the real data as simulated data being measured from updated pipelines become available from both the Roman and LSST surveys in the future. In this study, we select seed galaxies based on the SVM boundaries in Figure~\ref{fig:blue-red-gal-spe}. Figure~\ref{fig:c_vs_z_plot_seed} shows color versus true redshift (spec-z) distribution for the galaxy sample used in this study after making the magnitude cut. The orange points highlight the final seed galaxies selected for the red-sequence template calibration. 

\subsection{Roman Grism red galaxy efficiency curve} \label{sec: Grism efficiency}

In this study, we have assumed that spectroscopic redshift of all the seed galaxies will be sourced from the Roman high latitude spectroscopy survey (HLSS). And we use the true redshift recorded in the truth catalog as the spectroscopic redshift. However, not all galaxies selected based on the method outlined in Section~\ref{sec: seed galaxy selection} will have reliable spectroscopic counterparts from the Roman HLSS. Thus, it is important to account for the realistic performance of the Roman HLSS in estimating spec-z for LRGs. Given the absence of previous studies focused on assessing this aspect, we have developed a Roman grism image simulation to extract and analyze this necessary information.

We provide a brief summary here and direct readers to~\citet{Guo24} for more comprehensive details. Using simulated direct images from~\citet{Troxel23} as the reference, we simulate Roman grism images based on the Roman grism's instrumental properties, expected background levels, and the Roman reference survey design. Following this, we extract spectra of red, quiescent galaxies from the simulated grism images and perform redshift fitting. The simulation and proceeding analysis are all done by using the Grism Redshift and Line Analysis software, \verb|Grizli|\footnote{\url{https://github.com/gbrammer/grizli}}~\citep{Grizli}. Our analysis is focused on evaluating the redshift recovery efficiency, defined as the fraction of red, quiescent galaxies with high-confidence spectroscopic redshift estimates with respect to the total population of red galaxies in the simulation, across a redshift range of 1 to 2. 

\begin{figure}[t]
    \centering
    \plotone{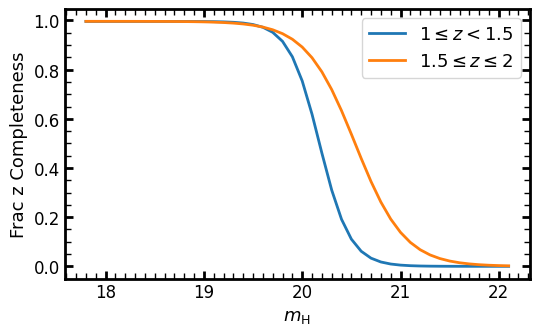}
    \caption{Roman grism efficiency curve for red, quiescent galaxies derived in~\citet{Guo24}.}
    \label{fig:grism-eff-curve}
\end{figure}

The final data product is the red galaxy efficiency curve, which reflects the fraction of red galaxies with high signal-to-noise ratio (SNR $>$ 5) and have reliable spec-z measurements ($(z_\mathrm{spec} - z_\mathrm{truth})/(1 + z_\mathrm{truth}) \leq 0.01$), as a function of $H$ band magnitude. This curve, derived based on the Roman reference survey strategy, acts as a selection function indicating the probability of a red galaxy at a specific redshift and with a specific magnitude being identified and having redshift measured accurately.  Additionally, we observed that, with similar brightness, red galaxies at higher redshifts ($z \geq 1.5$) typically exhibit more accurate redshift fitting results compared to those at lower redshifts. This improved accuracy is attributed to the Roman grism's wavelength coverage of $1\mu m - 2\mu m$, which aligns with the redshifted $4000\mathrm{\AA}$ break in galaxies within the redshift range of $1.5\leq z \leq 4$. The presence of the break in the extracted spectra enhances the precision of redshift measurements. Thus, we apply two sets of efficiency curves, as shown in Figure~\ref{fig:grism-eff-curve}, to select seed galaxy in two redshift intervals, $1 \leq z \leq 1.5$ and $1.5\leq z \leq 2$ respectively from the seed galaxy sample.

\subsection{Scaling of overlap region} \label{sec: sacaling of overlap region}

One major issue we identified after incorporating the red galaxy efficiency curve into the seed galaxy selection pipeline is that there are not enough seed galaxies at higher redshift that can be used for calibration. This is because higher-redshift galaxies tend to be fainter and the efficiency curve drops significantly at fainter magnitudes.  However, the simulation we work with only covers a $20\ \mathrm{deg}^2$ overlapping field of Roman and LSST, and the current Roman reference survey assumes a full overlap region of $2000\ \mathrm{deg}^2$. To fully utilize the overlap, we assume that the current simulated $20\ \mathrm{deg}^2$ field is representative of the entire survey area. We then repeatedly apply the efficiency curve to the initial seed galaxy sample as the assumption of the coverage of the overlapping region increases. For example, when considering a $40\ \mathrm{deg}^2$ region, the selection function is applied twice, and the samples from both runs are combined to form the final seed galaxy sample. This inevitably introduces duplicate galaxies—particularly when extended to larger regions—resulting in a sample increasingly dominated by duplicates, especially at lower redshifts ($1 < z < 1.5$), where most bright galaxies reside and the efficiency curve approaches unity for such galaxies. This may lead to over-simplified results. However, this approach still provides valuable insights into how improved statistics from a larger overlap region can allow us to identify LRGs and measure redshift at higher redshifts.

To address the limitations of the scaling approach, we adopt a bootstrap-based method as a more robust cross-check of our results. Rather than artificially increasing the survey area by duplicating galaxies, this approach captures the statistical variation in the selection process driven by instrumental noise.
Specifically, we generate multiple realizations of the grism selection efficiency by resampling the detector-level outputs from the Roman grism simulation. Since the simulation is performed across 18 detectors, each providing an independent noise realization, we construct new realizations by sampling these detectors with replacement and recomputing the efficiency curve as a function of galaxy properties (e.g., $H$-band magnitude).
For each bootstrap iteration, the resulting efficiency curve is applied to the same $20\ \mathrm{deg}^2$ seed galaxy sample, followed by an independent fit of the red-sequence template. This produces a distribution of red-sequence parameter estimates, from which the final parameters are obtained by averaging over all realizations.
This bootstrap-based framework preserves the underlying galaxy population while incorporating realistic variations in selection efficiency, and therefore serves as a statistically well-motivated cross-check of the results obtained from the simple scaling approach. We find that the results from this bootstrap-based method are consistent with those derived from the scaling approach.

The results presented in the following subsections are based on a full overlap of $2000\ \mathrm{deg}^2$ for illustrative purposes. We will discuss results based on different assumptions regarding the overlap area in Section~\ref{sec: maxz_vs_overlaparea}.

\subsection{Red-sequence template calibration} \label{sec: rs-caliberation}

\begin{figure}
    \centering
    \includegraphics[width=0.9\linewidth]{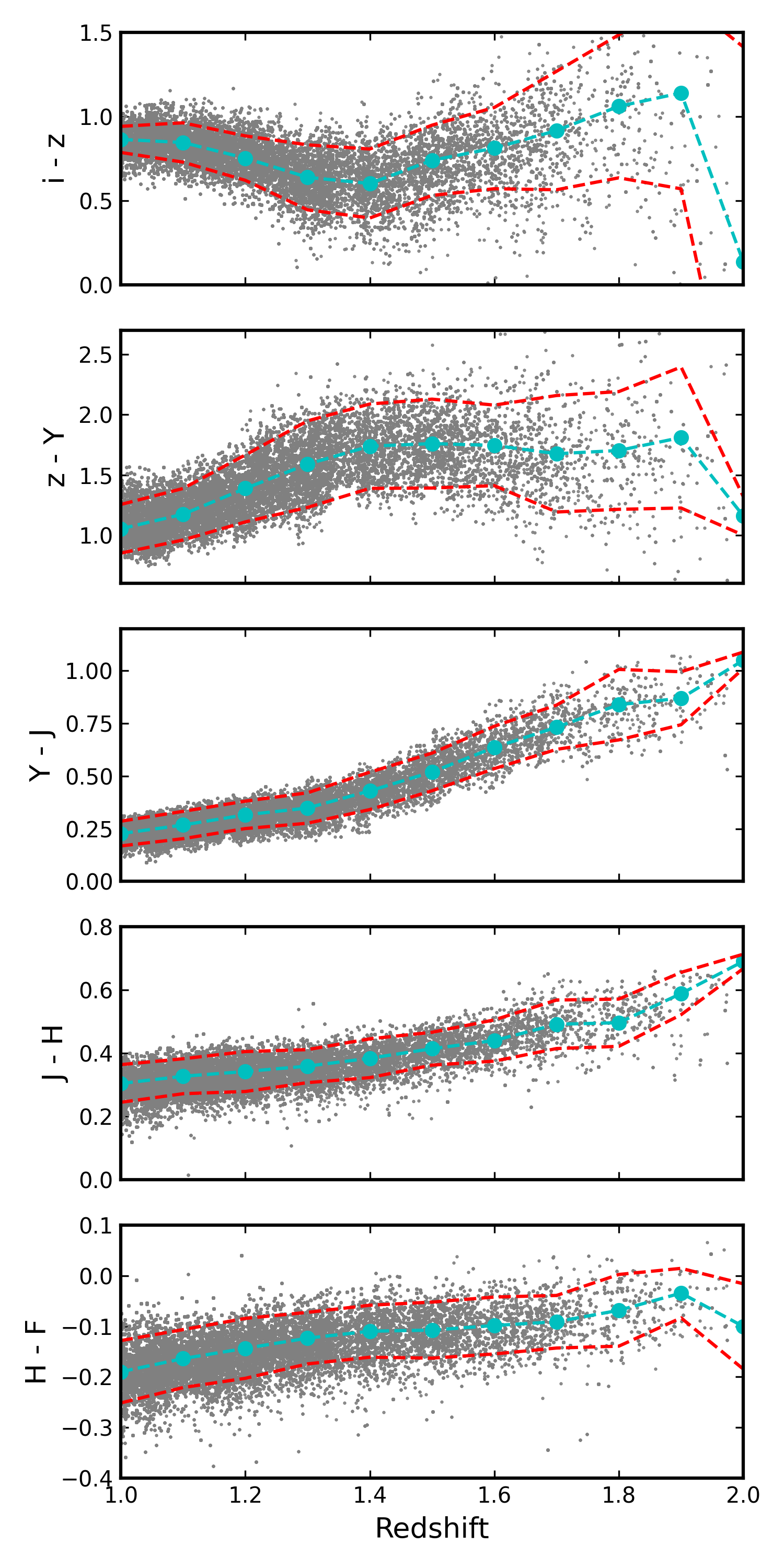}
    \caption{Color ($i - z,\ z - Y,\ Y - J,\ J - H,\ H - F$) as a function of redshift for the selected seed galaxies. The cyan points indicate the $a(z)$ values at the spline node positions, and the cyan, dashed lines are the spline interpolation. The red, dashed lines indicates the $3\sigma_{\mathrm{int}}$ range. Conversely, the larger number of outliers in the five colors above reflects the fact that the photometric errors are larger than the intrinsic width of the red-sequence.}
    \label{fig:c_z_caliberation}
\end{figure}

\begin{figure*}
    \centering
    \epsscale{1.2}
    \plotone{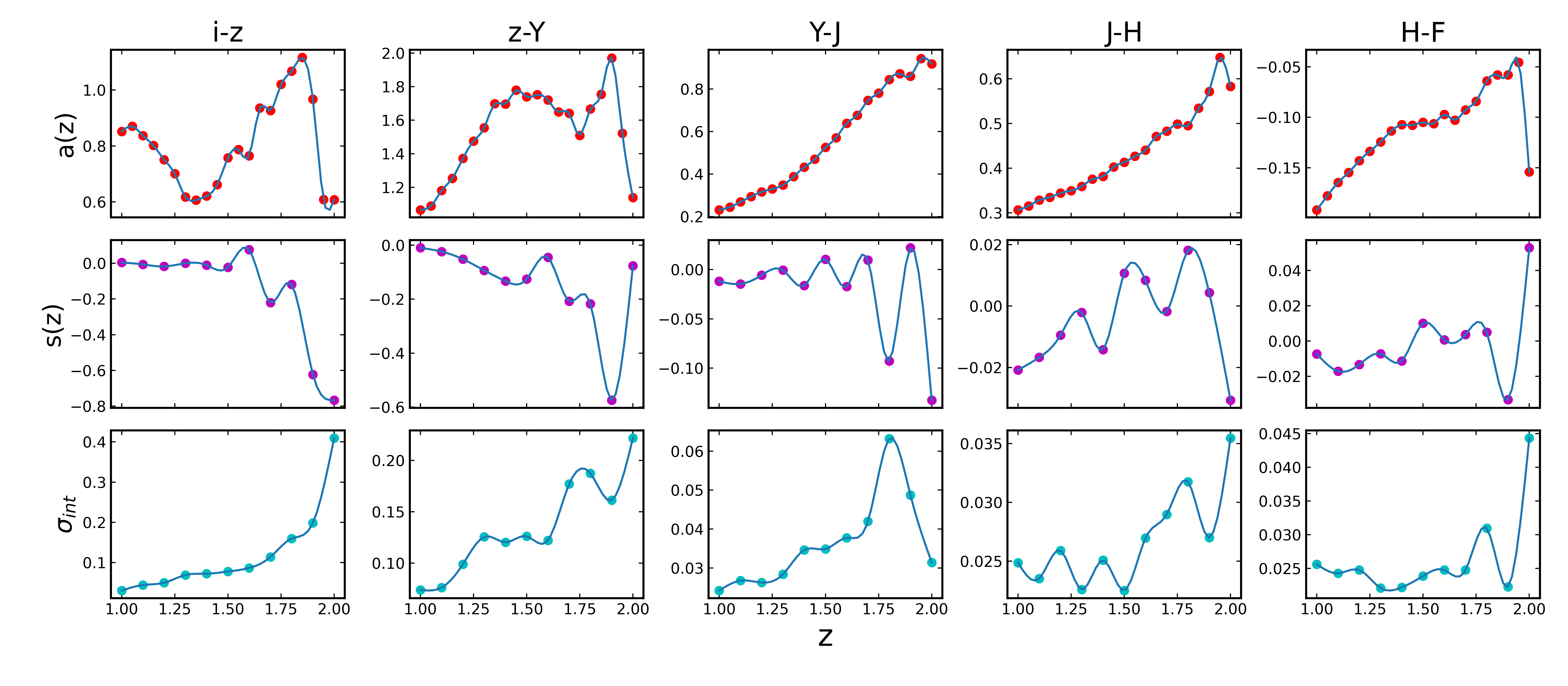}
    \caption{Caliberated Red-sequence parameters as a function of redshift from $z = 1$ to $2$. From left to right, each panel displays the trends for different colors: $i - z,\ z - Y,\ Y - J,\ J - H,\ H - F$. The top panels show the mean color ($C(z)$) across redshift at each redshift node (red points), the middle panels show the slope of the red-sequence template ($S(z)$) as each redshift node (purple points), and the bottom panels show the intrinsic width ($\sigma_{\text{int}}$) of the fitted color model at each redshift node (cyan points). The solid curves show the fitted spline interpolation function.}
    \label{fig:rs-parameters}
\end{figure*}

The red-sequence template is defined by four parameters: intercept $\mathbf{a}(z)$, slope, $\mathbf{s}(z)$, intrinsic scatter, $\mathbf{C_{\mathrm{int}}}(z)$, and reference magnitude, $m_{\mathrm{ref}}(z)$ in Equation~\ref{eq:rs-model}. This section outlines the calibration process for these parameters by using the seed galaxy samples described in previous sections, which closely follows the methodology presented in~\citet{Rykoff14} .

\textit{Step 1: Pivot magnitude estimations:} 
The first step of calibration is to estimate pivot magnitudes $m_{\mathrm{ref}}(z)$, which is estimated using cubic spline interpolation based on a set of $m_{\mathrm{ref}}(z_i)$ values. These are calculated as the median magnitude of the seed galaxies at spline nodes $i$, which are uniformly distributed from $z = 1$ to $z = 2$.

\textit{Step 2: Perform initial estimates of mean-color redshift relation and its width to refine seed galaxy sample: } Having defined the pivot magnitudes as function of redshift, The second step is to perform an initial estimate of the mean color-redshift relation, $\Tilde{c}(z)$ for each color. This is achieved by minimizing the objective function:
\begin{equation} \label{eq: mean-c-min}
    O = \sum_i\left|c_i - \Tilde{c}(z_i)\right|
\end{equation}
where $i$ sums over all seed galaxies used for calibration, and $\Tilde{c}(z)$ is determined through spline interpolation, with the spline node values—minimized in Equation~\ref{eq: mean-c-min}—defined on a redshift grid spaced at intervals of 0.1. The minimization is performed by using the \verb|scipy.optimize.minimize| function with the \verb|'L-BFGS-B'| method. Next the initial estimation of mean-color is used to calculated an initial estimation of width of the color-redshift relation for each color, $\Tilde{\sigma}_{\mathrm{int}}(z)$. This is done by minimizing the objective function,
\begin{equation}
    O = \sum_i |c_i - \Tilde{c}(z_i) - MAD|
\end{equation}
where $MAD$ stands for the median absolute deviation about the median and it is related to $\Tilde{\sigma}_{\mathrm{int}}(z)$ by:
\begin{equation}
    \Tilde{\sigma}_{\mathrm{int}}(z) = 1.4826 \times MAD
\end{equation}
Next, we refine the seed sample by selecting galaxies within $ \Tilde{\sigma}_{\mathrm{int}}(z)$ of the median color. This selection criterion ensures that the calibration leverages the core of the red galaxy distribution, minimizing potential bias from bluer galaxies.

\textit{Step 3: Calibration of intercept, slope and intrinsic scatter:} The updated seed galaxy sample and pivot defined, we now turn to calibrate the intercept, $\mathbf{a}(z)$, slope $\mathbf{s}(z)$ and intrinsic scatter $\mathbf{\sigma_{\mathrm{int}}}$ of the red-sequence relation in Equation~\ref{eq:rs-model}. Similarly, cublic spline interpolation functions are used to parameterize smoothly evolving functions of redshift. In this study, we choose $0.05$, $0.1$ and $0.1$ as the node spacing for intercept, slope and intrinsic scatter respectively. This is achieved by minimizing the negative log-likelihood function for each color respectively:
\begin{align}
    O &= -\sum_i\log{\Tilde{\mathcal{L}}_i} \\
      &= -\sum_i\log\left[\frac{1}{\sqrt{2\pi\sigma_{\mathrm{tot}}^2}}\exp{\left(-\frac{1}{2}\left(\frac{c_i - c(z,m)}{\sigma_{\mathrm{tot}}}\right)^2\right)}\right]
\end{align}
where $c(z,m)$ is given by Equation~\ref{eq:rs-model}, $c_i$ is the color for each seed galaxy, and $\sigma_{\mathrm{tot}} = (\sigma_{\mathrm{int}}^2(z) + \sigma_{c,i}^2)^{1/2}$ with $\sigma_{c,i}^2$ being the color error for each individual galaxy. 

The calibration of the intercept, $\mathbf{a}(z)$, slope, $\mathbf{s}(z)$, and intrinsic scatter of each color, $\mathbf{\sigma_{\mathrm{int}}}(z)$ that constitutes the diagonal element of $\mathbf{C_{\mathrm{int}}}(z)$, follows a sequential approach. Initially, all parameters are estimated with initial guess values. We begin with fitting the intercept first, then update this estimate to refine the slope, and finally adjust the intrinsic scatter. Once all parameters have been individually refined, a final joint calibration is conducted with these updated values serving as the initial guesses. This approach ensures each parameter is optimally tuned, enhancing the overall accuracy and reliability of the red-sequence model. To visualize the calibration results, we present Figure~\ref{fig:c_z_caliberation}, which shows the color-redshift relations for the seed galaxies selected based on a $2000\ \deg^2$ overlap region. The cyan points represent the mean color calibrated at each spline node, while the red short-dashed line indicates the $3\sigma_\mathrm{int}$ region. Figure~\ref{fig:rs-parameters} displays the final calibration of all the red-sequence template parameters as a function of redshift for the five colors used in this study. Beyond $z = 1.75$, the spline node parameters of slope and intrinsic scatter for the calibrated template are poorly constrained,  This is likely due to the lack of seed galaxies at redshifts $z > 1.75$, as shown in Figure~\ref{fig:c_vs_z_plot_seed} and~\ref{fig:c_z_caliberation}. Additionally, the $z-Y$ color exhibits unusually high intrinsic scatter compared to other colors. This may be due to differences in how photometry is measured in the DC2 catalog and the Roman simulated data, as discussed in Section~\ref{sec:data}.

\subsection{Selection criteria} \label{sec: red galaxy selection}

The selection of LRG candidates in this study applies two criteria following~\citet{Rozo16}. The first criterion is a luminosity threshold to ensure that the galaxies are sufficiently luminous. Previous research created two samples based on two luminosity thresholds: a \verb|dense| (hereafter referred to as \verb|highdens|) sample with a luminosity ratio $l > 0.5$ (see Eq.~\ref{eq:lum-ratio}), and a \verb|luminous| (hereafter \verb|highlum|) sample with $l > 1$.

The second criterion is a $\chi^2$ cut on the  $\chi^2_{\mathrm{red}}$ (see Equation~\ref{eq: chi2_red}) computed at the best-fit redshift, $\widetilde{z_\mathrm{red}}$, which ensures that the selected galaxies closely align with the calibrated red-sequence. Previous studies~\citep{Rozo16,Vakili19,Vakili23} implemented a variable $\chi^2_{\mathrm{red}}(z)$ that adjusts at different redshifts to maintain a constant co-moving density of the samples. In such cases, the derived $\chi_{\mathrm{red}}^2(z)$ is relatively small (e.g. $2-3$), resulting in a sample that tightly follows the calibrated red-sequence template. However, it has been noted that while a variable $\chi_{\mathrm{red}}^2$ cut tends to minimize outliers, it also increases correlations with systematic errors in photometry. 
Specifically, ~\citet{Pandey22} has found that the DES-Y3 RedMaGiC sample derived in this way leads to inconsistencies in the galaxy clustering and galaxy-galaxy lensing results. They proposed that this behavior might be related to unknown systematics that have not been corrected for, such as a color-dependent photometric issue in the DES data. \citet{Pandey22} further found that this issue can be resolved by loosening the $\chi^2$ threshold via defining a broader, constant $\chi_{\mathrm{red}}^2 = 8$ cut to the sample.  In response to these issues, our study follows~\citet{Pandey22} and  applies a simple, constant cut of $\chi^2_{\mathrm{red}} \leq 8$ to select LRGs, avoiding the complexities and potential biases introduced by a variable threshold.

\begin{figure*}[t]
    \centering
    \epsscale{1.15}
    \plottwo{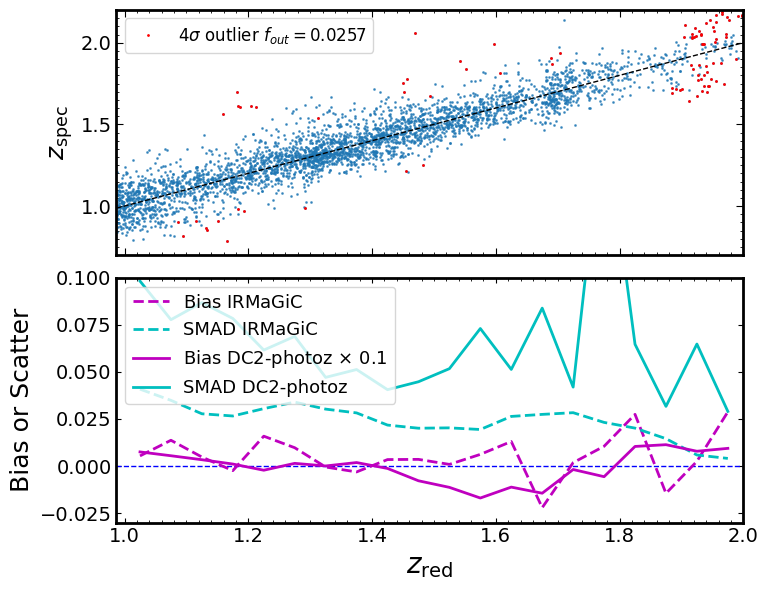}{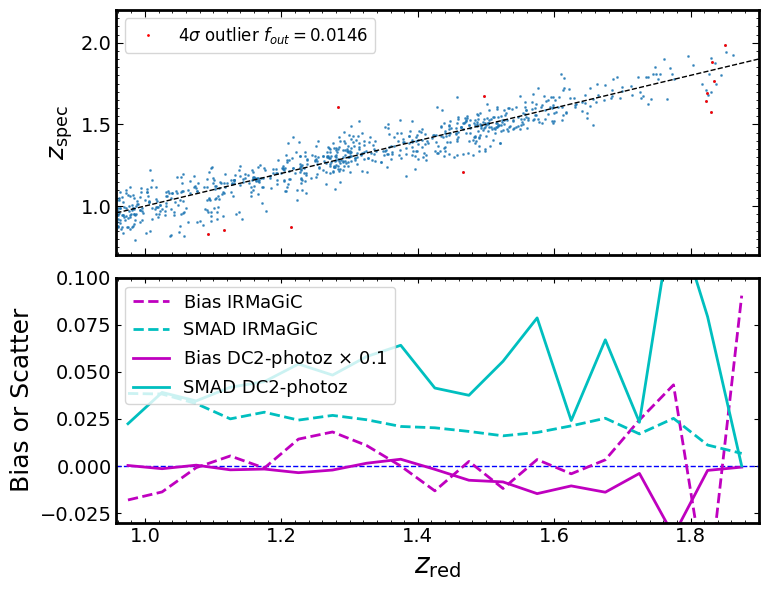}
    \caption{Top panels display the estimated red-sequence photometric redshift $z_{\mathrm{red}}$ for galaxies in the \textbf{dense}, $L > 0.5L_*$, (left) and \textbf{highlum},  $L > L_*$, (right) samples versus their true redshift $z_{\mathrm{spec}}$. The red dots show the $4\sigma$ outlier. The bottom panel shows the bias and scatter: the dash (solid) purple line represents the mean bias $z_{\mathrm{red}} - z_{\mathrm{spec}}$ calculated across redshift bins for IRMaGiC redshift (DC2-photoz estimated using BPz), and the cyan dashed (solid) line depicts the scatter or NMAD as defined in Equation~\ref{eq:NMAD}, across the $z_{\mathrm{red}}$ for IRMaGiC redshift (DC2-photoz).}
    \label{fig:highden-lum-results}
\end{figure*}

\subsection{Photo-z afterburner} \label{sec: afterburner}
When a subset of galaxies in the selected IRMaGiC sample has spectroscopic redshift measurements, these galaxies can be used to calibrate the median redshift offset $\Delta_z (\widetilde{z_{\mathrm{red}}}) = z_{\mathrm{spec}} - \widetilde{z_{\mathrm{red}}}$. And thus obtain the final, refined photometric redshift:
\begin{equation}
    z_{\mathrm{red}} =  \widetilde{z_{\mathrm{red}}} + \Delta _z(\widetilde{z_{\mathrm{red}}})
\end{equation}
And $\Delta z(\widetilde{z_{\mathrm{red}}})$ parameterized using a spline interpolation, with spline parameters optimized by minimizing the following objective function.:
\begin{equation}
    O = \sum_i \left|z_{\mathrm{spec,i}} - \widetilde{z_{\mathrm{red,i}}} \right|
\end{equation}
To obtain such a sample for afterburner calibration, we applied the Roman grism efficiency curves introduced in Section~\ref{sec: Grism efficiency} to IRMaGiC sample, simulating a scenario in which galaxies with spectroscopic redshifts are available within the sample. The true redshift of selected galaxies are then used for calibration. In this study, we assume that $30\%$ of the galaxies in the highlum/highdens sample will have spectroscopy redshift. Additionally, we can also refine the estimated photometric redshift error, $\sigma_\mathrm{red}$. Assuming the refined error $\sigma_{\mathrm{refine}}$ is given by $\sigma_{\mathrm{refine}} = r(z_{\mathrm{red}})\sigma_\mathrm{red}$, and the $r(z_{\mathrm{red}})$ parameter is again characterized by a spline interpolation with best-fit parameters minimizing the objective function:
\begin{equation}
    O = \sum_i \left| 1.4826 \left| z_{\textrm{red}, i} - z_{\textrm{spec}, i} \right| - r(z_{\textrm{red}, i}) \sigma_{{\mathrm{red}}, i} \right|
\end{equation}
We refer the reader to~\citet{Rozo16} for more details on this calibration process.
After the afterburner step, we recomputed the luminosity ratio $l$ and $\chi^2_{\mathrm{red}}$, as discussed in Section~\ref{sec: red galaxy selection}, for every galaxy in the highdens/highlum sample with refined photometric redshifts and errors.

\section{Results} \label{sec:results}

\subsection{Photo-z performance}\label{sec: photo-z performance}

As discussed in Section~\ref{sec: red galaxy selection}, we consider two sets of IRMaGiC sample. The \verb|highdens| sample with galaxies brighter than $0.5L_*$ and the \verb|highlum| sample with galaxies brighter than $L_*$. Additionally, unless otherwise stated, all the results presented in this section are based on the assumption of a $2000\deg^2$ full overlap between LSST and Roman. We note that our sample does not have constant comoving density for reasons discussed in Section~\ref{sec: red galaxy selection}. Following~\citet{Rozo16,Vakili19}, we assess the performance of the IRMaGiC photo-z of the selected galaxies by calculating two quantities in bins of the afterburner calibrated redshift $z_\mathrm{red}$. The photometric redshift bias, defined as the median offset of $\delta_z = z_\mathrm{spec} - z_\mathrm{red}$, and scatter, defined as  Normalised Absolute Median Deviation (NMAD)
\begin{equation} \label{eq:NMAD}
    \mathrm{NMAD} = 1.4826\times\mathrm{median(\left|\Delta_z\right|)}
\end{equation}
where $\Delta_z = (z_\mathrm{spec} - z_\mathrm{red})/(1+z_\mathrm{spec})$.

Figure~\ref{fig:highden-lum-results} shows the performance of the estimated IRMaGiC photo-z for the \verb|highdens| sample (left-hand panel) and the \verb|highlum| sample (right-hand panel). The purple and cyan dashed lines in the bottom panels present the bias and scatter, binned by $z_\mathrm{red}$. Several key points merit attention. First, the \verb|highdens| sample experiences a significant number of $5\sigma$ outliers beyond $z_{\mathrm{red}} \approx 1.9$. This can be attributed to the absence of seed galaxies necessary for a well-calibrated red-sequence template in this redshift range as discussed in section~\ref{sec: rs-caliberation}. As indicated in the legend of the top panels, the overall $5\sigma$ fraction of the \verb|highdens| sample from $z_\mathrm{red} = 1$ to $2$ is approximately $4\%$. However, restricting this to $z_\mathrm{red} = 1$ to $1.9$, the $5\sigma$ outlier rate drops to only $0.09\%$. 

Second, the \verb|highlum| sample contains no galaxies beyond $z_\mathrm{red} > 1.8$, which is consistent with the tendency of galaxies to appear fainter at higher redshifts and the luminosity threshold of the \verb|highlum| sample, which targets exceptionally bright galaxies. We report the mean bias and scatter for the \verb|highdens| and \verb|highlum| samples as ($3.2\times10^{-3}$, 0.026) and ($3.3\times10^{-3}$, 0.022), calculated within the redshift ranges of $z_\mathrm{red} = 1 - 1.9$ and $z_\mathrm{red} = 1 - 1.8$ respectively. Focusing on the scatter within the \verb|highdens| sample, we observe that the three highest scatter values occur at $z_\mathrm{red} \approx 1, 1.3,$ and $1.8$, respectively. This pattern can be explained by the transition of the $4000\mathrm{\AA}$ break across different filters: it moves from the LSST i-band to the LSST z-band at $z \approx 1$, from the LSST z-band to the Roman F106 band at $z \approx 1.3$, and from the Roman F106 band to the Roman F129 band at $z \approx 1.8$ as illustrated in Figure~\ref{fig:LSST+Roman_bands}.

\subsection{Comparing with DC2 photo-z} \label{sec: compare with DC2 redmagic}
We now evaluate the performance of IRMaGiC photo-z alongside other photo-z algorithms within the DC2 catalog, which we term as DC2-photoz in this paper, for each galaxy. These estimates are derived from the Bayesian Photometric Redshift (BPz) code, which leverages Bayesian probability to enhance redshift estimate accuracy. This method incorporates prior probabilities and applies Bayesian marginalization, integrating critical data such as redshift distributions and galaxy type mixes into the estimation process, thus enhancing precision. We refer readers to~\citet{Ben´ıtez00, Schmidt2020} for further details. Notably, the LSST-DESC Redshift Assessment Infrastructure Layer (RAIL)~\citep{Schmidt-Rail}, a framework for photometric redshift estimation and analysis for LSST-DESC, implements a similar version of the BPz code. Therefore, the recorded photo-z values provide an excellent preview of the quality of LSST photo-zs. In the bottom panels of Figure~\ref{fig:highden-lum-results}, we show the bias (purple, solid line) and scatter (cyan, solid line) of DC2-photozs for the \verb|highdens| and \verb|highlum| samples. Note that we scale the bias of DC2-photozs down by a factor of 10 for easier comparison with the IRMaGiC photo-z performance. The mean bias and scatter of the DC2-photoz are ($-$0.016, 0.069) and ($-$0.057, 0.05) for the \verb|highdens| and \verb|highlum| samples respectively. Comparing values of IRMaGiC photo-z in Section~\ref{sec: photo-z performance}, it is clear that IRMaGiC photo-zs outperform DC2-photoz. This suggests that the inclusion of additional data from the Roman Space Telescope could further refine the photometric redshift estimates of LSST galaxies beyond those achievable using LSST data alone.

\subsection{Comparing with low-z LRG samples}
\begin{figure*}[t]
    \centering
    \includegraphics[width=0.49\linewidth]{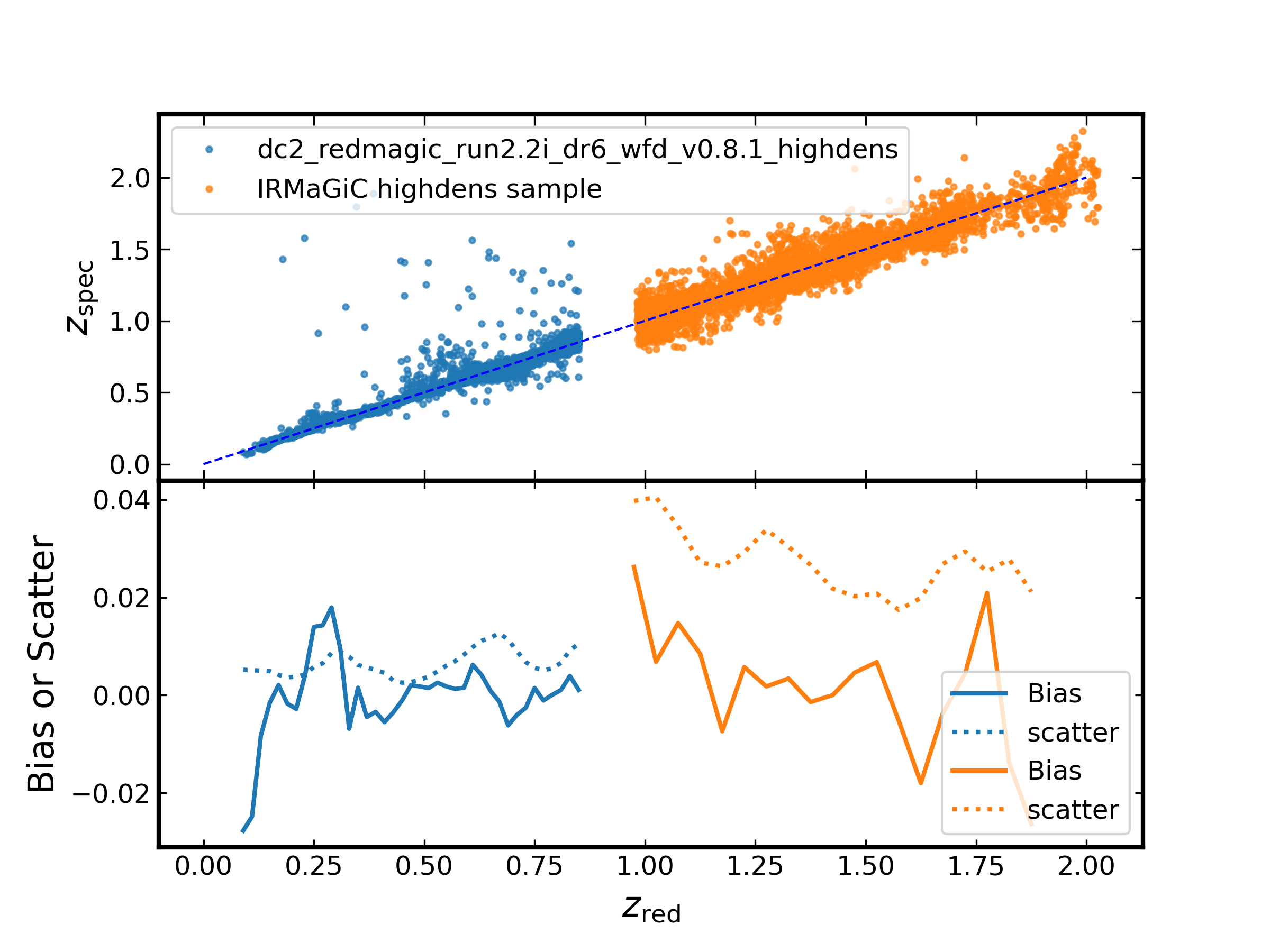}
    \includegraphics[width=0.49\linewidth]{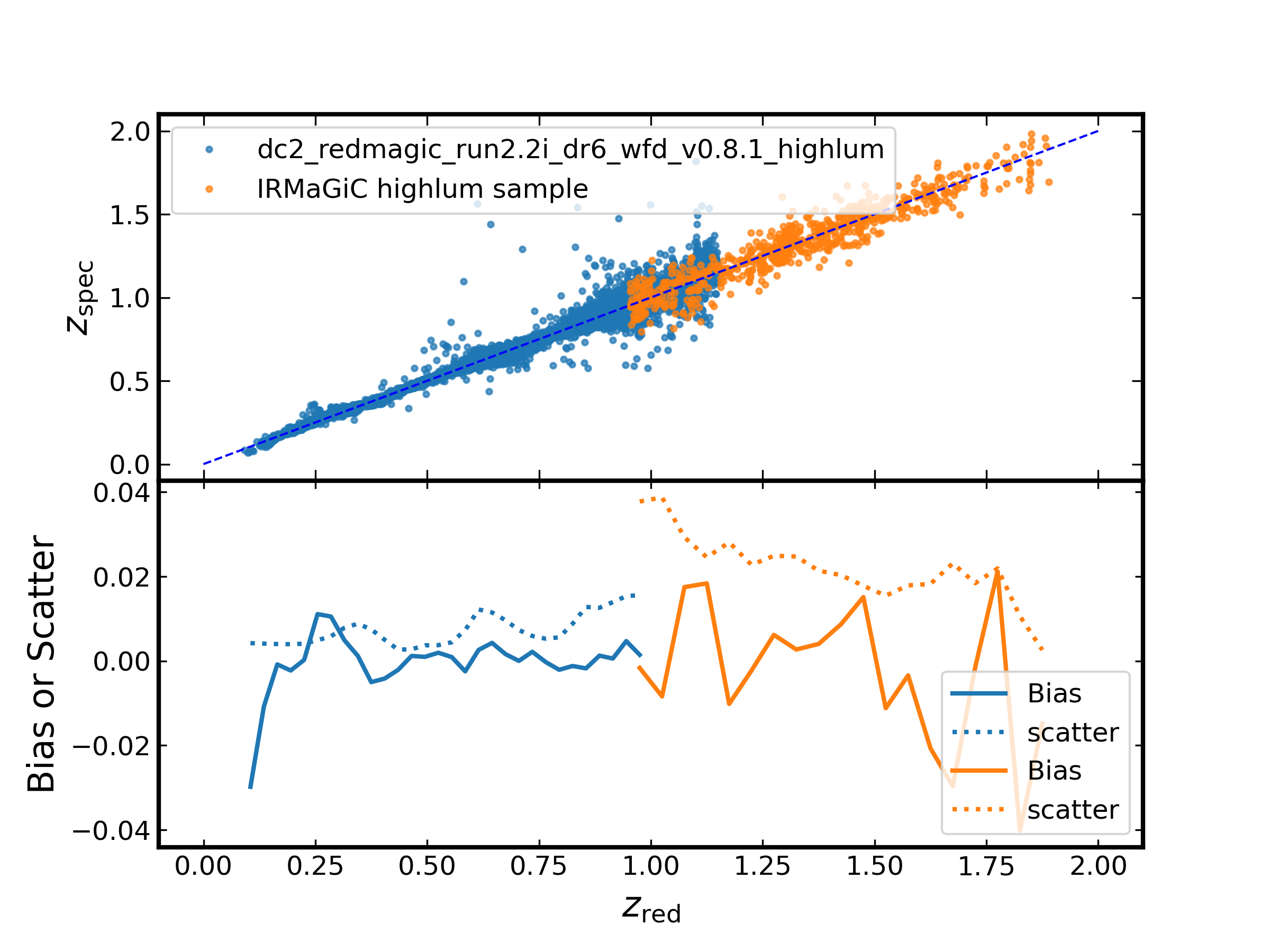}
    \caption{Combination of the DC2 RedMaGiC sample and LRG samples presented in this study. The left plot shows the combined highdens sample, the right plot shows the combined highlum sample. Top panels show the estimated redshift from the two samples versus true redshift $z_{\mathrm{spec}}$. The bottom panels show the bias (solid line) and Scatter/NMAD (dotted line) of the two samples.}
    \label{fig:combined-LRG-sample-compare}
\end{figure*}

The original RedMaPPer and RedMaGiC algorithms have been applied to LSST cosmoDC2/DC2 galaxies using only LSST photometry to generate LRG samples in the redshift range of $0 < z < 1$. The cosmoDC2/DC2 RedMaGiC sample applies constant $\chi^2$ cut for LRG selection, rather than a variable $\chi^2$, for the same reasons discussed in Section~\ref{sec: red galaxy selection}. In this section, we compare our IRMaGiC samples to the LSST DC2 RedMaGiC samples. Figure~\ref{fig:combined-LRG-sample-compare} illustrates the redshift performance comparison for the \verb|highdens|/\verb|highlum| samples from the LSST DC2 and IRMaGiC. As mentioned in Section~\ref{sec: photo-z performance}, our LRG samples exhibit greater scatter compared to other low-z RedMaGiC or RedMaGiC-like samples, this is also clearly shown in the bottom panel of Figure~\ref{fig:combined-LRG-sample-compare}. 

This increased scatter could be due to three factors. First, the underlying truth magnitudes for each galaxy differ between cosmoDC2, which the DC2 simulation is based on, and the Roman truth catalog. In cosmoDC2, the \verb|mag_norm| values used to scale the input galaxy SEDs vary across different bandpasses~\citep{LSSTDC221}, while the Roman simulation uses a constant \verb|mag_norm| to scale the SEDs across all Roman bands~\citep{Troxel23}. This difference in truth magnitude calculation may contribute to the increased scatter. Second, as noted in Section~\ref{sec:data}, the photometry in the two catalogs is measured differently, which could also contribute to the increased scatter. Third, red and blue galaxy populations tend to be blended more at higher redshifts, which broadens the intrinsic scatter of the red-sequence template.

To explore this further, we rerun the IRMaGiC algorithm on the truth catalog for the LSST-Roman joint galaxy sample. We calibrate a red-sequence model based on ground-truth information and compare the redshift performance to the cosmoDC2 RedMaGiC samples. Comparing results derived based on galaxy truth photometry allows us to rule out the impact of measured photometry. Our comparison shows that, despite differing assumptions about the scaling of galaxy SEDs and our sample being at higher redshifts,  the scatters are comparable in this case. Further details can be found in Appendix~\ref{appendix}. We therefore conclude that the increased scatter observed here is likely driven by the current measurements in Roman magnitude, which are measured using a generic SourceExtractor photometry pipeline (See Section~\ref{sec: Roman data description}). In contrast, we expect the final Roman data products to be generated with a survey-optimized photometric pipeline, yielding improved calibration and more accurate photometry measurements.

Moreover, Figure~\ref{fig:combined-LRG-sample-compare} demonstrates the significant extension of existing LRG samples to higher redshifts by combining data from LSST and Roman. Expanding the LRG sample at higher redshifts is crucial for improving future cosmological analyses. For example, in the HSC Y3 weak lensing study, the tomographic redshift distribution was inferred by combining individual photo-z posteriors with cross-correlation measurements of the HSC CAMIRA LRG sample. However, the redshift distribution could not be fully calibrated due to the absence of a reliable calibration sample of CAMIRA LRGs at $z \geq 1$~\citep{Rau23,Miyatake23}. The extended LRG sample from LSST and Roman will help resolve similar issues in the future, enabling more accurate redshift calibration at higher redshifts and ultimately leading to tighter cosmological constraints.

\subsection{Redshift coverage as a function of overlapping area} \label{sec: maxz_vs_overlaparea}

\begin{table*}[t]
\centering
\caption{Photometric redshift performance of the highdens and highlum IRMaGiC galaxy samples for different overlapping areas between LSST and Roman. $z_{\mathrm{max}}$ is the maximum redshift range of the sample.\label{tbl:pverlap-area-summary}}
\begin{tabular}{c|cccc|cccc}
\hline
\multirow{2}{*}{Overlapping area ($\deg^2$)} & \multicolumn{4}{c|}{Highdens} & \multicolumn{4}{c}{Highlum} \\
\cline{2-9}
 & $z_\mathrm{max}$ & Bias (\%) & Scatter (\%) & Outlier (\%) & $z_\mathrm{max}$ & Bias (\%) & Scatter (\%) & Outlier (\%) \\
\hline
20   & 1.70 & 0.21 & 3.05 & 1.30 & 1.68 & 0.70 & 2.48 & 1.41 \\
40   & 1.72 & 0.31 & 2.60 & 1.70 & 1.72 & 0.50 & 2.50 & 0.71 \\
100  & 1.82 & 0.36 & 2.56 & 1.07 & 1.76 & 0.75 & 2.39 & 1.53 \\
200  & 1.88 & 0.13 & 2.70 & 1.36 & 1.80 & 0.39 & 2.36 & 0.43 \\
500  & 1.92 & 0.26 & 2.64 & 1.68 & 1.79 & 0.79 & 2.33 & 1.16 \\
1000 & 1.92 & 0.15 & 2.65 & 1.06 & 1.80 & 0.16 & 2.40 & 0.57 \\
2000 & 1.93 & 0.14 & 2.63 & 1.01 & 1.81 & 0.04 & 2.30 & 1.20 \\
\hline
\end{tabular}
\end{table*}

The analysis thus far has assumed a complete overlap of $2000\deg^2$ between the LSST and the Roman surveys. In this section, we expand our examination to assess the impact of varying the size of the overlap region on our results. Following the methodology outlined in Section~\ref{sec: sacaling of overlap region}, in addition to the initial $2000\deg^2$ template, we calibrated additional red-sequence templates for overlap regions of $20\deg^2$, $40\deg^2$, $100\deg^2$, $500\deg^2$, and $1000\deg^2$. For each newly calibrated template, we run on the same galaxy sample to calculate the mean bias, mean scatter, outlier rate, and to determine the maximum redshift coverage. The final maximum redshift for each scenario is determined based on visual inspections, where a significant rise in the frequency of $5\sigma$ outliers indicates the limit of reliable redshift. The comprehensive results of these analyses are summarized in Table~\ref{tbl:pverlap-area-summary}, detailing the effects of each overlap scenario on IRMaGiC photo-z performance. The most substantial improvements in redshift coverage occur from $40\deg^2$ ($z_{\text{max}} = 1.72/1.72$) to $100\deg^2$ ($z_{\text{max}} = 1.82/1.76$) and from $100\deg^2$ to $200\deg^2$ ($z_{\text{max}} = 1.88/1.8$). Beyond a $500\deg^2$ overlap, further improvements in redshift coverage are minimal. Therefore, an overlap region of at least $200\deg^2$ is necessary for IRMaGiC to achieve reliable redshift measurements up to $z \approx 1.9$ and $1.8$ for each sample respectively.

We note, however, that these results may also be affected by the limitations inherent in the scaling procedure described in Section~\ref{sec: sacaling of overlap region}. As larger and more realistic simulated survey areas, or eventually real data, become available, this limitation can be mitigated, and the quantitative results presented here can be refined in future studies.

\section{Summary} \label{sec:summary}

In this paper, we develop a RedMaGiC-like algorithm to identify luminous red galaxies (LRGs) in the redshift range of $1 \leq z \leq 2$ using simulated data from two Stage IV dark energy surveys: the Vera C. Rubin Observatory's Legacy Survey of Space and Time (LSST) and the Nancy Grace Roman Space Telescope's High Latitude Wide-Area Survey. We utilize the complementary wavelength coverage of LSST's optical bands and Roman HLIS's infrared bands to extend the red-sequence calibration to higher redshifts.

Our method involves cross-matching simulated photometric data from LSST and Roman HLIS and using the Roman High Latitude Spectroscopy Survey (HLSS) to supply spectroscopically confirmed red galaxies for the red-sequence template calibration. We present a new strategy for selecting high-redshift red galaxy candidates and calibrating the red-sequence template, which allows for accurate photometric redshift estimates. The performance of our algorithm is evaluated using two luminosity thresholds for LRG selection.

The results show that our RedMaGiC-like algorithm achieves robust photometric redshift accuracy, outperforming existing photometric redshift estimation methods in the LSST DC2 catalog. Additionally, we explore the impact of the overlap region size between LSST and Roman on redshift coverage and find that an overlap of at least 200 deg² is necessary to reliably extend redshift measurements to $z \approx 1.9$. These findings highlight the potential of joint LSST-Roman observations to enhance LRG selection and photometric redshift estimation for cosmological studies.

\begin{acknowledgements}
This paper has undergone internal review by the LSST Dark Energy Science Collaboration. The internal reviewers were Irene Moskowitz and Anna Porredon.

ZG and CW were supported by Department of Energy, grant DE-SC0010007. 

The DESC acknowledges ongoing support from the Institut National de Physique Nucl\'eaire et de Physique des Particules in France; the Science \& Technology Facilities Council in the United Kingdom; and theDepartment of Energy, the National Science Foundation, and the LSST Corporation in the United States.  DESC uses resources of the IN2P3 Computing Center (CC-IN2P3--Lyon/Villeurbanne - France) funded by the Centre National de la Recherche Scientifique; the National Energy Research Scientific Computing Center, a DOE Office of Science User Facility supported by the Office of Science of the U.S.\ Department ofEnergy under Contract No.\ DE-AC02-05CH11231; STFC DiRAC HPC Facilities, funded by UK BEIS National E-infrastructure capital grants; and the UK particle physics grid, supported by the GridPP Collaboration.  This work was performed in part under DOE Contract DE-AC02-76SF00515.

The contributions from the primary authors are as follows. ZG analysed data, and wrote the paper. CW supervised the project and analysis. E.R. contributed expertise and code related to the Red-Sequence fitting algorithm used in RedMapper and RedMaGiC.
\end{acknowledgements}

\bibliographystyle{aasjournal}

\bibliography{sample631}

@article{Oguri18b,
    author = {Oguri, Masamune and Lin, Yen-Ting and Lin, Sheng-Chieh and Nishizawa, Atsushi J and More, Anupreeta and More, Surhud and Hsieh, Bau-Ching and Medezinski, Elinor and Miyatake, Hironao and Jian, Hung-Yu and Lin, Lihwai and Takada, Masahiro and Okabe, Nobuhiro and Speagle, Joshua S and Coupon, Jean and Leauthaud, Alexie and Lupton, Robert H and Miyazaki, Satoshi and Price, Paul A and Tanaka, Masayuki and Chiu, I-Non and Komiyama, Yutaka and Okura, Yuki and Tanaka, Manobu M and Usuda, Tomonori},
    title = "{An optically-selected cluster catalog at redshift 0.1 \\&lt; z \\&lt; 1.1 from the Hyper Suprime-Cam Subaru Strategic Program S16A data}",
    journal = {Publications of the Astronomical Society of Japan},
    volume = {70},
    number = {SP1},
    pages = {S20},
    year = {2017},
    month = {06},
    issn = {0004-6264},
    doi = {10.1093/pasj/psx042},
    url = {https://doi.org/10.1093/pasj/psx042},
    eprint = {https://academic.oup.com/pasj/article-pdf/70/SP1/S20/54675166/pasj\_70\_sp1\_s20.pdf},
}

@article{Vakili19,
    author = {Vakili, Mohammadjavad and Bilicki, Maciej and Hoekstra, Henk and Chisari, Nora Elisa and Brown, Michael J I and Georgiou, Christos and Kannawadi, Arun and Kuijken, Konrad and Wright, Angus H},
    title = "{Luminous red galaxies in the Kilo-Degree Survey: selection with broad-band photometry and weak lensing measurements}",
    journal = {Monthly Notices of the Royal Astronomical Society},
    volume = {487},
    number = {3},
    pages = {3715-3733},
    year = {2019},
    month = {05},
    issn = {0035-8711},
    doi = {10.1093/mnras/stz1249},
    url = {https://doi.org/10.1093/mnras/stz1249},
    eprint = {https://academic.oup.com/mnras/article-pdf/487/3/3715/28845648/stz1249.pdf},
}

@ARTICLE{Eisenstein01,
       author = {{Eisenstein}, Daniel J. and {Annis}, James and {Gunn}, James E. and {Szalay}, Alexander S. and {Connolly}, Andrew J. and {Nichol}, R.~C. and {Bahcall}, Neta A. and {Bernardi}, Mariangela and {Burles}, Scott and {Castander}, Francisco J. and {Fukugita}, Masataka and {Hogg}, David W. and {Ivezi{\'c}}, {\v{Z}}eljko and {Knapp}, G.~R. and {Lupton}, Robert H. and {Narayanan}, Vijay and {Postman}, Marc and {Reichart}, Daniel E. and {Richmond}, Michael and {Schneider}, Donald P. and {Schlegel}, David J. and {Strauss}, Michael A. and {SubbaRao}, Mark and {Tucker}, Douglas L. and {Vanden Berk}, Daniel and {Vogeley}, Michael S. and {Weinberg}, David H. and {Yanny}, Brian},
        title = "{Spectroscopic Target Selection for the Sloan Digital Sky Survey: The Luminous Red Galaxy Sample}",
      journal = {\aj},
         year = 2001,
        month = nov,
       volume = {122},
       number = {5},
        pages = {2267-2280},
          doi = {10.1086/323717},
archivePrefix = {arXiv},
       eprint = {astro-ph/0108153},
 primaryClass = {astro-ph},
       adsurl = {https://ui.adsabs.harvard.edu/abs/2001AJ....122.2267E}
}

@ARTICLE{Zhou23,
       author = {{Zhou}, Rongpu and {Dey}, Biprateep and {Newman}, Jeffrey A. and {Eisenstein}, Daniel J. and {Dawson}, K. and {Bailey}, S. and {Berti}, A. and {Guy}, J. and {Lan}, Ting-Wen and {Zou}, H. and {Aguilar}, J. and {Ahlen}, S. and {Alam}, Shadab and {Brooks}, D. and {de la Macorra}, A. and {Dey}, A. and {Dhungana}, G. and {Fanning}, K. and {Font-Ribera}, A. and {Gontcho}, S. Gontcho A. and {Honscheid}, K. and {Ishak}, Mustapha and {Kisner}, T. and {Kov{\'a}cs}, A. and {Kremin}, A. and {Landriau}, M. and {Levi}, Michael E. and {Magneville}, C. and {Manera}, Marc and {Martini}, P. and {Meisner}, Aaron M. and {Miquel}, R. and {Moustakas}, J. and {Myers}, Adam D. and {Nie}, Jundan and {Palanque-Delabrouille}, N. and {Percival}, W.~J. and {Poppett}, C. and {Prada}, F. and {Raichoor}, A. and {Ross}, A.~J. and {Schlafly}, E. and {Schlegel}, D. and {Schubnell}, M. and {Tarl{\'e}}, Gregory and {Weaver}, B.~A. and {Wechsler}, R.~H. and {Y{\'e}che}, Christophe and {Zhou}, Zhimin},
        title = "{Target Selection and Validation of DESI Luminous Red Galaxies}",
      journal = {\aj},
         year = 2023,
        month = feb,
       volume = {165},
       number = {2},
          eid = {58},
        pages = {58},
          doi = {10.3847/1538-3881/aca5fb},
archivePrefix = {arXiv},
       eprint = {2208.08515},
 primaryClass = {astro-ph.CO},
       adsurl = {https://ui.adsabs.harvard.edu/abs/2023AJ....165...58Z}
}

@ARTICLE{Bisigello20,
       author = {{Bisigello}, L. and {Kuchner}, U. and {Conselice}, C.~J. and {Andreon}, S. and {Bolzonella}, M. and {Duc}, P. -A. and {Garilli}, B. and {Humphrey}, A. and {Maraston}, C. and {Moresco}, M. and {Pozzetti}, L. and {Tortora}, C. and {Zamorani}, G. and {Auricchio}, N. and {Brinchmann}, J. and {Capobianco}, V. and {Carretero}, J. and {Castander}, F.~J. and {Castellano}, M. and {Cavuoti}, S. and {Cimatti}, A. and {Cledassou}, R. and {Congedo}, G. and {Conversi}, L. and {Corcione}, L. and {Cropper}, M.~S. and {Dusini}, S. and {Frailis}, M. and {Franceschi}, E. and {Franzetti}, P. and {Fumana}, M. and {Hormuth}, F. and {Israel}, H. and {Jahnke}, K. and {Kermiche}, S. and {Kitching}, T. and {Kohley}, R. and {Kubik}, B. and {Kunz}, M. and {Le F{\`e}vre}, O. and {Ligori}, S. and {Lilje}, P.~B. and {Lloro}, I. and {Maiorano}, E. and {Marggraf}, O. and {Massey}, R. and {Masters}, D.~C. and {Mei}, S. and {Mellier}, Y. and {Meylan}, G. and {Padilla}, C. and {Paltani}, S. and {Pasian}, F. and {Pettorino}, V. and {Pires}, S. and {Polenta}, G. and {Poncet}, M. and {Raison}, F. and {Rhodes}, J. and {Roncarelli}, M. and {Rossetti}, E. and {Saglia}, R. and {Sauvage}, M. and {Schneider}, P. and {Secroun}, A. and {Serrano}, S. and {Sureau}, F. and {Taylor}, A.~N. and {Tereno}, I. and {Toledo-Moreo}, R. and {Valenziano}, L. and {Wang}, Y. and {Wetzstein}, M. and {Zoubian}, J.},
        title = "{Euclid: the selection of quiescent and star-forming galaxies using observed colours}",
      journal = {\mnras},
         year = 2020,
        month = may,
       volume = {494},
       number = {2},
        pages = {2337-2354},
          doi = {10.1093/mnras/staa885},
archivePrefix = {arXiv},
       eprint = {2003.07367},
 primaryClass = {astro-ph.GA},
       adsurl = {https://ui.adsabs.harvard.edu/abs/2020MNRAS.494.2337B}
}

@ARTICLE{Rozo16,
       author = {{Rozo}, E. and {Rykoff}, E.~S. and {Abate}, A. and {Bonnett}, C. and {Crocce}, M. and {Davis}, C. and {Hoyle}, B. and {Leistedt}, B. and {Peiris}, H.~V. and {Wechsler}, R.~H. and {Abbott}, T. and {Abdalla}, F.~B. and {Banerji}, M. and {Bauer}, A.~H. and {Benoit-L{\'e}vy}, A. and {Bernstein}, G.~M. and {Bertin}, E. and {Brooks}, D. and {Buckley-Geer}, E. and {Burke}, D.~L. and {Capozzi}, D. and {Rosell}, A. Carnero and {Carollo}, D. and {Kind}, M. Carrasco and {Carretero}, J. and {Castander}, F.~J. and {Childress}, M.~J. and {Cunha}, C.~E. and {D'Andrea}, C.~B. and {Davis}, T. and {DePoy}, D.~L. and {Desai}, S. and {Diehl}, H.~T. and {Dietrich}, J.~P. and {Doel}, P. and {Eifler}, T.~F. and {Evrard}, A.~E. and {Neto}, A. Fausti and {Flaugher}, B. and {Fosalba}, P. and {Frieman}, J. and {Gaztanaga}, E. and {Gerdes}, D.~W. and {Glazebrook}, K. and {Gruen}, D. and {Gruendl}, R.~A. and {Honscheid}, K. and {James}, D.~J. and {Jarvis}, M. and {Kim}, A.~G. and {Kuehn}, K. and {Kuropatkin}, N. and {Lahav}, O. and {Lidman}, C. and {Lima}, M. and {Maia}, M.~A.~G. and {March}, M. and {Martini}, P. and {Melchior}, P. and {Miller}, C.~J. and {Miquel}, R. and {Mohr}, J.~J. and {Nichol}, R.~C. and {Nord}, B. and {O'Neill}, C.~R. and {Ogando}, R. and {Plazas}, A.~A. and {Romer}, A.~K. and {Roodman}, A. and {Sako}, M. and {Sanchez}, E. and {Santiago}, B. and {Schubnell}, M. and {Sevilla-Noarbe}, I. and {Smith}, R.~C. and {Soares-Santos}, M. and {Sobreira}, F. and {Suchyta}, E. and {Swanson}, M.~E.~C. and {Thaler}, J. and {Thomas}, D. and {Uddin}, S. and {Vikram}, V. and {Walker}, A.~R. and {Wester}, W. and {Zhang}, Y. and {da Costa}, L.~N.},
        title = "{redMaGiC: selecting luminous red galaxies from the DES Science Verification data}",
      journal = {\mnras},
         year = 2016,
        month = sep,
       volume = {461},
       number = {2},
        pages = {1431-1450},
          doi = {10.1093/mnras/stw1281},
archivePrefix = {arXiv},
       eprint = {1507.05460},
 primaryClass = {astro-ph.IM},
       adsurl = {https://ui.adsabs.harvard.edu/abs/2016MNRAS.461.1431R}
}

@ARTICLE{Rykoff14,
       author = {{Rykoff}, E.~S. and {Rozo}, E. and {Busha}, M.~T. and {Cunha}, C.~E. and {Finoguenov}, A. and {Evrard}, A. and {Hao}, J. and {Koester}, B.~P. and {Leauthaud}, A. and {Nord}, B. and {Pierre}, M. and {Reddick}, R. and {Sadibekova}, T. and {Sheldon}, E.~S. and {Wechsler}, R.~H.},
        title = "{redMaPPer. I. Algorithm and SDSS DR8 Catalog}",
      journal = {\apj},
         year = 2014,
        month = apr,
       volume = {785},
       number = {2},
          eid = {104},
        pages = {104},
          doi = {10.1088/0004-637X/785/2/104},
archivePrefix = {arXiv},
       eprint = {1303.3562},
 primaryClass = {astro-ph.CO},
       adsurl = {https://ui.adsabs.harvard.edu/abs/2014ApJ...785..104R}
}

@ARTICLE{Crocce19,
       author = {{Crocce}, M. and {Ross}, A.~J. and {Sevilla-Noarbe}, I. and {Gaztanaga}, E. and {Elvin-Poole}, J. and {Avila}, S. and {Alarcon}, A. and {Chan}, K.~C. and {Banik}, N. and {Carretero}, J. and {Sanchez}, E. and {Hartley}, W.~G. and {S{\'a}nchez}, C. and {Giannantonio}, T. and {Rosenfeld}, R. and {Salvador}, A.~I. and {Garcia-Fernandez}, M. and {Garc{\'\i}a-Bellido}, J. and {Abbott}, T.~M.~C. and {Abdalla}, F.~B. and {Allam}, S. and {Annis}, J. and {Bechtol}, K. and {Benoit-L{\'e}vy}, A. and {Bernstein}, G.~M. and {Bernstein}, R.~A. and {Bertin}, E. and {Brooks}, D. and {Buckley-Geer}, E. and {Carnero Rosell}, A. and {Carrasco Kind}, M. and {Castander}, F.~J. and {Cawthon}, R. and {Cunha}, C.~E. and {D'Andrea}, C.~B. and {da Costa}, L.~N. and {Davis}, C. and {De Vicente}, J. and {Desai}, S. and {Diehl}, H.~T. and {Doel}, P. and {Drlica-Wagner}, A. and {Eifler}, T.~F. and {Fosalba}, P. and {Frieman}, J. and {Garc{\'\i}a-Bellido}, J. and {Gerdes}, D.~W. and {Gruen}, D. and {Gruendl}, R.~A. and {Gschwend}, J. and {Gutierrez}, G. and {Hollowood}, D. and {Honscheid}, K. and {Jain}, B. and {James}, D.~J. and {Krause}, E. and {Kuehn}, K. and {Kuhlmann}, S. and {Kuropatkin}, N. and {Lahav}, O. and {Lima}, M. and {Maia}, M.~A.~G. and {Marshall}, J.~L. and {Martini}, P. and {Menanteau}, F. and {Miller}, C.~J. and {Miquel}, R. and {Nichol}, R.~C. and {Percival}, W.~J. and {Plazas}, A.~A. and {Sako}, M. and {Scarpine}, V. and {Schindler}, R. and {Scolnic}, D. and {Sheldon}, E. and {Smith}, M. and {Smith}, R.~C. and {Soares-Santos}, M. and {Sobreira}, F. and {Suchyta}, E. and {Swanson}, M.~E.~C. and {Tarle}, G. and {Thomas}, D. and {Tucker}, D.~L. and {Vikram}, V. and {Walker}, A.~R. and {Yanny}, B. and {Zhang}, Y. and {Dark Energy Survey Collaboration}},
        title = "{Dark Energy Survey year 1 results: galaxy sample for BAO measurement}",
      journal = {\mnras},
         year = 2019,
        month = jan,
       volume = {482},
       number = {2},
        pages = {2807-2822},
          doi = {10.1093/mnras/sty2522},
archivePrefix = {arXiv},
       eprint = {1712.06211},
 primaryClass = {astro-ph.CO},
       adsurl = {https://ui.adsabs.harvard.edu/abs/2019MNRAS.482.2807C}
}

@ARTICLE{Troxel23,
       author = {{Troxel}, M.~A. and {Lin}, C. and {Park}, A. and {Hirata}, C. and {Mandelbaum}, R. and {Jarvis}, M. and {Choi}, A. and {Givans}, J. and {Higgins}, M. and {Sanchez}, B. and {Yamamoto}, M. and {Awan}, H. and {Chiang}, J. and {Dor{\'e}}, O. and {Walter}, C.~W. and {Zhang}, T. and {Cohen-Tanugi}, J. and {Gawiser}, E. and {Hearin}, A. and {Heitmann}, K. and {Ishak}, M. and {Kovacs}, E. and {Mao}, Y. -Y. and {Wood-Vasey}, M. and {Becker}, Matt and {Meyers}, Josh and {Melchior}, Peter and {LSST Dark Energy Science Collaboration}},
        title = "{A joint Roman Space Telescope and Rubin Observatory synthetic wide-field imaging survey}",
      journal = {\mnras},
         year = 2023,
        month = jun,
       volume = {522},
       number = {2},
        pages = {2801-2820},
          doi = {10.1093/mnras/stad664},
archivePrefix = {arXiv},
       eprint = {2209.06829},
 primaryClass = {astro-ph.IM},
       adsurl = {https://ui.adsabs.harvard.edu/abs/2023MNRAS.522.2801T}
}

@software{Grizli,
       author = {{Brammer}, Gabe},
        title = "{Grizli: Grism redshift and line analysis software}",
 howpublished = {Astrophysics Source Code Library, record ascl:1905.001},
         year = 2019,
        month = may,
          eid = {ascl:1905.001},
       adsurl = {https://ui.adsabs.harvard.edu/abs/2019ascl.soft05001B}
}

@ARTICLE{Ben´ıtez00,
       author = {{Ben{\'\i}tez}, Narciso},
        title = "{Bayesian Photometric Redshift Estimation}",
      journal = {\apj},
         year = 2000,
        month = jun,
       volume = {536},
       number = {2},
        pages = {571-583},
          doi = {10.1086/308947},
archivePrefix = {arXiv},
       eprint = {astro-ph/9811189},
 primaryClass = {astro-ph},
       adsurl = {https://ui.adsabs.harvard.edu/abs/2000ApJ...536..571B}
}

@ARTICLE{Schmidt2020,
       author = {{Schmidt}, S.~J. and {Malz}, A.~I. and {Soo}, J.~Y.~H. and {Almosallam}, I.~A. and {Brescia}, M. and {Cavuoti}, S. and {Cohen-Tanugi}, J. and {Connolly}, A.~J. and {DeRose}, J. and {Freeman}, P.~E. and {Graham}, M.~L. and {Iyer}, K.~G. and {Jarvis}, M.~J. and {Kalmbach}, J.~B. and {Kovacs}, E. and {Lee}, A.~B. and {Longo}, G. and {Morrison}, C.~B. and {Newman}, J.~A. and {Nourbakhsh}, E. and {Nuss}, E. and {Pospisil}, T. and {Tranin}, H. and {Wechsler}, R.~H. and {Zhou}, R. and {Izbicki}, R. and {LSST Dark Energy Science Collaboration}},
        title = "{Evaluation of probabilistic photometric redshift estimation approaches for The Rubin Observatory Legacy Survey of Space and Time (LSST)}",
      journal = {\mnras},
         year = 2020,
        month = dec,
       volume = {499},
       number = {2},
        pages = {1587-1606},
          doi = {10.1093/mnras/staa2799},
archivePrefix = {arXiv},
       eprint = {2001.03621},
 primaryClass = {astro-ph.CO},
       adsurl = {https://ui.adsabs.harvard.edu/abs/2020MNRAS.499.1587S}
}

@software{Schmidt-Rail,
  author       = {Sam Schmidt and
                  Julia Gschwend and
                  John Franklin Crenshaw and
                  Zi'ang Yan and
                  Eric Charles and
                  Alex Malz and
                  Shahab Joudaki and
                  Olivia R. Lynn and
                  Luca Tortorelli and
                  hangqianjun and
                  joezuntz and
                  imoskowitz and
                  Bryce Kalmbach and
                  jlvdb and
                  sylvielsstfr and
                  Johann Cohen-Tanugi and
                  Josue De Santiago and
                  Drew Oldag and
                  Melissa DeLucchi and
                  sjs86 and
                  vladislav doster and
                  Francois Lanusse and
                  Heather Kelly},
  title        = {LSSTDESC/RAIL: v0.98.5},
  month        = may,
  year         = 2023,
  publisher    = {Zenodo},
  version      = {v0.98.5},
  doi          = {10.5281/zenodo.7927358},
  url          = {https://doi.org/10.5281/zenodo.7927358}
}

@ARTICLE{Guo24,
       author = {{Guo}, Zhiyuan and {Joshi}, Bhavin and {Walter}, C.~W. and {Troxel}, M.~A.},
        title = "{Simulating Continuum-based Redshift Measurement in the Roman's High Latitude Spectroscopic Survey}",
      journal = {\aj},
         year = 2025,
        month = jun,
       volume = {169},
       number = {6},
          eid = {320},
        pages = {320},
          doi = {10.3847/1538-3881/adcd6b},
archivePrefix = {arXiv},
       eprint = {2411.08035},
 primaryClass = {astro-ph.GA},
       adsurl = {https://ui.adsabs.harvard.edu/abs/2025AJ....169..320G}
}

@ARTICLE{Albrecht06,
       author = {{Albrecht}, Andreas and {Bernstein}, Gary and {Cahn}, Robert and {Freedman}, Wendy L. and {Hewitt}, Jacqueline and {Hu}, Wayne and {Huth}, John and {Kamionkowski}, Marc and {Kolb}, Edward W. and {Knox}, Lloyd and {Mather}, John C. and {Staggs}, Suzanne and {Suntzeff}, Nicholas B.},
        title = "{Report of the Dark Energy Task Force}",
      journal = {arXiv e-prints},
         year = 2006,
        month = sep,
          eid = {astro-ph/0609591},
        pages = {astro-ph/0609591},
          doi = {10.48550/arXiv.astro-ph/0609591},
archivePrefix = {arXiv},
       eprint = {astro-ph/0609591},
 primaryClass = {astro-ph},
       adsurl = {https://ui.adsabs.harvard.edu/abs/2006astro.ph..9591A}
}

@ARTICLE{Ivezic19,
       author = {{Ivezi{\'c}}, {\v{Z}}eljko and {Kahn}, Steven M. and {Tyson}, J. Anthony and {Abel}, Bob and {Acosta}, Emily and {Allsman}, Robyn and {Alonso}, David and {AlSayyad}, Yusra and {Anderson}, Scott F. and {Andrew}, John and {Angel}, James Roger P. and {Angeli}, George Z. and {Ansari}, Reza and {Antilogus}, Pierre and {Araujo}, Constanza and {Armstrong}, Robert and {Arndt}, Kirk T. and {Astier}, Pierre and {Aubourg}, {\'E}ric and {Auza}, Nicole and {Axelrod}, Tim S. and {Bard}, Deborah J. and {Barr}, Jeff D. and {Barrau}, Aurelian and {Bartlett}, James G. and {Bauer}, Amanda E. and {Bauman}, Brian J. and {Baumont}, Sylvain and {Bechtol}, Ellen and {Bechtol}, Keith and {Becker}, Andrew C. and {Becla}, Jacek and {Beldica}, Cristina and {Bellavia}, Steve and {Bianco}, Federica B. and {Biswas}, Rahul and {Blanc}, Guillaume and {Blazek}, Jonathan and {Blandford}, Roger D. and {Bloom}, Josh S. and {Bogart}, Joanne and {Bond}, Tim W. and {Booth}, Michael T. and {Borgland}, Anders W. and {Borne}, Kirk and {Bosch}, James F. and {Boutigny}, Dominique and {Brackett}, Craig A. and {Bradshaw}, Andrew and {Brandt}, William Nielsen and {Brown}, Michael E. and {Bullock}, James S. and {Burchat}, Patricia and {Burke}, David L. and {Cagnoli}, Gianpietro and {Calabrese}, Daniel and {Callahan}, Shawn and {Callen}, Alice L. and {Carlin}, Jeffrey L. and {Carlson}, Erin L. and {Chandrasekharan}, Srinivasan and {Charles-Emerson}, Glenaver and {Chesley}, Steve and {Cheu}, Elliott C. and {Chiang}, Hsin-Fang and {Chiang}, James and {Chirino}, Carol and {Chow}, Derek and {Ciardi}, David R. and {Claver}, Charles F. and {Cohen-Tanugi}, Johann and {Cockrum}, Joseph J. and {Coles}, Rebecca and {Connolly}, Andrew J. and {Cook}, Kem H. and {Cooray}, Asantha and {Covey}, Kevin R. and {Cribbs}, Chris and {Cui}, Wei and {Cutri}, Roc and {Daly}, Philip N. and {Daniel}, Scott F. and {Daruich}, Felipe and {Daubard}, Guillaume and {Daues}, Greg and {Dawson}, William and {Delgado}, Francisco and {Dellapenna}, Alfred and {de Peyster}, Robert and {de Val-Borro}, Miguel and {Digel}, Seth W. and {Doherty}, Peter and {Dubois}, Richard and {Dubois-Felsmann}, Gregory P. and {Durech}, Josef and {Economou}, Frossie and {Eifler}, Tim and {Eracleous}, Michael and {Emmons}, Benjamin L. and {Fausti Neto}, Angelo and {Ferguson}, Henry and {Figueroa}, Enrique and {Fisher-Levine}, Merlin and {Focke}, Warren and {Foss}, Michael D. and {Frank}, James and {Freemon}, Michael D. and {Gangler}, Emmanuel and {Gawiser}, Eric and {Geary}, John C. and {Gee}, Perry and {Geha}, Marla and {Gessner}, Charles J.~B. and {Gibson}, Robert R. and {Gilmore}, D. Kirk and {Glanzman}, Thomas and {Glick}, William and {Goldina}, Tatiana and {Goldstein}, Daniel A. and {Goodenow}, Iain and {Graham}, Melissa L. and {Gressler}, William J. and {Gris}, Philippe and {Guy}, Leanne P. and {Guyonnet}, Augustin and {Haller}, Gunther and {Harris}, Ron and {Hascall}, Patrick A. and {Haupt}, Justine and {Hernandez}, Fabio and {Herrmann}, Sven and {Hileman}, Edward and {Hoblitt}, Joshua and {Hodgson}, John A. and {Hogan}, Craig and {Howard}, James D. and {Huang}, Dajun and {Huffer}, Michael E. and {Ingraham}, Patrick and {Innes}, Walter R. and {Jacoby}, Suzanne H. and {Jain}, Bhuvnesh and {Jammes}, Fabrice and {Jee}, M. James and {Jenness}, Tim and {Jernigan}, Garrett and {Jevremovi{\'c}}, Darko and {Johns}, Kenneth and {Johnson}, Anthony S. and {Johnson}, Margaret W.~G. and {Jones}, R. Lynne and {Juramy-Gilles}, Claire and {Juri{\'c}}, Mario and {Kalirai}, Jason S. and {Kallivayalil}, Nitya J. and {Kalmbach}, Bryce and {Kantor}, Jeffrey P. and {Karst}, Pierre and {Kasliwal}, Mansi M. and {Kelly}, Heather and {Kessler}, Richard and {Kinnison}, Veronica and {Kirkby}, David and {Knox}, Lloyd and {Kotov}, Ivan V. and {Krabbendam}, Victor L. and {Krughoff}, K. Simon and {Kub{\'a}nek}, Petr and {Kuczewski}, John and {Kulkarni}, Shri and {Ku}, John and {Kurita}, Nadine R. and {Lage}, Craig S. and {Lambert}, Ron and {Lange}, Travis and {Langton}, J. Brian and {Le Guillou}, Laurent and {Levine}, Deborah and {Liang}, Ming and {Lim}, Kian-Tat and {Lintott}, Chris J. and {Long}, Kevin E. and {Lopez}, Margaux and {Lotz}, Paul J. and {Lupton}, Robert H. and {Lust}, Nate B. and {MacArthur}, Lauren A. and {Mahabal}, Ashish and {Mandelbaum}, Rachel and {Markiewicz}, Thomas W. and {Marsh}, Darren S. and {Marshall}, Philip J. and {Marshall}, Stuart and {May}, Morgan and {McKercher}, Robert and {McQueen}, Michelle and {Meyers}, Joshua and {Migliore}, Myriam and {Miller}, Michelle and {Mills}, David J. and {Miraval}, Connor and {Moeyens}, Joachim and {Moolekamp}, Fred E. and {Monet}, David G. and {Moniez}, Marc and {Monkewitz}, Serge and {Montgomery}, Christopher and {Morrison}, Christopher B. and {Mueller}, Fritz and {Muller}, Gary P. and {Mu{\~n}oz Arancibia}, Freddy and {Neill}, Douglas R. and {Newbry}, Scott P. and {Nief}, Jean-Yves and {Nomerotski}, Andrei and {Nordby}, Martin and {O'Connor}, Paul and {Oliver}, John and {Olivier}, Scot S. and {Olsen}, Knut and {O'Mullane}, William and {Ortiz}, Sandra and {Osier}, Shawn and {Owen}, Russell E. and {Pain}, Reynald and {Palecek}, Paul E. and {Parejko}, John K. and {Parsons}, James B. and {Pease}, Nathan M. and {Peterson}, J. Matt and {Peterson}, John R. and {Petravick}, Donald L. and {Libby Petrick}, M.~E. and {Petry}, Cathy E. and {Pierfederici}, Francesco and {Pietrowicz}, Stephen and {Pike}, Rob and {Pinto}, Philip A. and {Plante}, Raymond and {Plate}, Stephen and {Plutchak}, Joel P. and {Price}, Paul A. and {Prouza}, Michael and {Radeka}, Veljko and {Rajagopal}, Jayadev and {Rasmussen}, Andrew P. and {Regnault}, Nicolas and {Reil}, Kevin A. and {Reiss}, David J. and {Reuter}, Michael A. and {Ridgway}, Stephen T. and {Riot}, Vincent J. and {Ritz}, Steve and {Robinson}, Sean and {Roby}, William and {Roodman}, Aaron and {Rosing}, Wayne and {Roucelle}, Cecille and {Rumore}, Matthew R. and {Russo}, Stefano and {Saha}, Abhijit and {Sassolas}, Benoit and {Schalk}, Terry L. and {Schellart}, Pim and {Schindler}, Rafe H. and {Schmidt}, Samuel and {Schneider}, Donald P. and {Schneider}, Michael D. and {Schoening}, William and {Schumacher}, German and {Schwamb}, Megan E. and {Sebag}, Jacques and {Selvy}, Brian and {Sembroski}, Glenn H. and {Seppala}, Lynn G. and {Serio}, Andrew and {Serrano}, Eduardo and {Shaw}, Richard A. and {Shipsey}, Ian and {Sick}, Jonathan and {Silvestri}, Nicole and {Slater}, Colin T. and {Smith}, J. Allyn and {Smith}, R. Chris and {Sobhani}, Shahram and {Soldahl}, Christine and {Storrie-Lombardi}, Lisa and {Stover}, Edward and {Strauss}, Michael A. and {Street}, Rachel A. and {Stubbs}, Christopher W. and {Sullivan}, Ian S. and {Sweeney}, Donald and {Swinbank}, John D. and {Szalay}, Alexander and {Takacs}, Peter and {Tether}, Stephen A. and {Thaler}, Jon J. and {Thayer}, John Gregg and {Thomas}, Sandrine and {Thornton}, Adam J. and {Thukral}, Vaikunth and {Tice}, Jeffrey and {Trilling}, David E. and {Turri}, Max and {Van Berg}, Richard and {Vanden Berk}, Daniel and {Vetter}, Kurt and {Virieux}, Francoise and {Vucina}, Tomislav and {Wahl}, William and {Walkowicz}, Lucianne and {Walsh}, Brian and {Walter}, Christopher W. and {Wang}, Daniel L. and {Wang}, Shin-Yawn and {Warner}, Michael and {Wiecha}, Oliver and {Willman}, Beth and {Winters}, Scott E. and {Wittman}, David and {Wolff}, Sidney C. and {Wood-Vasey}, W. Michael and {Wu}, Xiuqin and {Xin}, Bo and {Yoachim}, Peter and {Zhan}, Hu},
        title = "{LSST: From Science Drivers to Reference Design and Anticipated Data Products}",
      journal = {\apj},
         year = 2019,
        month = mar,
       volume = {873},
       number = {2},
          eid = {111},
        pages = {111},
          doi = {10.3847/1538-4357/ab042c},
archivePrefix = {arXiv},
       eprint = {0805.2366},
 primaryClass = {astro-ph},
       adsurl = {https://ui.adsabs.harvard.edu/abs/2019ApJ...873..111I}
}

@ARTICLE{Spergel15,
       author = {{Spergel}, D. and {Gehrels}, N. and {Baltay}, C. and {Bennett}, D. and {Breckinridge}, J. and {Donahue}, M. and {Dressler}, A. and {Gaudi}, B.~S. and {Greene}, T. and {Guyon}, O. and {Hirata}, C. and {Kalirai}, J. and {Kasdin}, N.~J. and {Macintosh}, B. and {Moos}, W. and {Perlmutter}, S. and {Postman}, M. and {Rauscher}, B. and {Rhodes}, J. and {Wang}, Y. and {Weinberg}, D. and {Benford}, D. and {Hudson}, M. and {Jeong}, W. -S. and {Mellier}, Y. and {Traub}, W. and {Yamada}, T. and {Capak}, P. and {Colbert}, J. and {Masters}, D. and {Penny}, M. and {Savransky}, D. and {Stern}, D. and {Zimmerman}, N. and {Barry}, R. and {Bartusek}, L. and {Carpenter}, K. and {Cheng}, E. and {Content}, D. and {Dekens}, F. and {Demers}, R. and {Grady}, K. and {Jackson}, C. and {Kuan}, G. and {Kruk}, J. and {Melton}, M. and {Nemati}, B. and {Parvin}, B. and {Poberezhskiy}, I. and {Peddie}, C. and {Ruffa}, J. and {Wallace}, J.~K. and {Whipple}, A. and {Wollack}, E. and {Zhao}, F.},
        title = "{Wide-Field InfrarRed Survey Telescope-Astrophysics Focused Telescope Assets WFIRST-AFTA 2015 Report}",
      journal = {arXiv e-prints},
         year = 2015,
        month = mar,
          eid = {arXiv:1503.03757},
        pages = {arXiv:1503.03757},
          doi = {10.48550/arXiv.1503.03757},
archivePrefix = {arXiv},
       eprint = {1503.03757},
 primaryClass = {astro-ph.IM},
       adsurl = {https://ui.adsabs.harvard.edu/abs/2015arXiv150303757S}
}

@ARTICLE{Akeson19,
       author = {{Akeson}, Rachel and {Armus}, Lee and {Bachelet}, Etienne and {Bailey}, Vanessa and {Bartusek}, Lisa and {Bellini}, Andrea and {Benford}, Dominic and {Bennett}, David and {Bhattacharya}, Aparna and {Bohlin}, Ralph and {Boyer}, Martha and {Bozza}, Valerio and {Bryden}, Geoffrey and {Calchi Novati}, Sebastiano and {Carpenter}, Kenneth and {Casertano}, Stefano and {Choi}, Ami and {Content}, David and {Dayal}, Pratika and {Dressler}, Alan and {Dor{\'e}}, Olivier and {Fall}, S. Michael and {Fan}, Xiaohui and {Fang}, Xiao and {Filippenko}, Alexei and {Finkelstein}, Steven and {Foley}, Ryan and {Furlanetto}, Steven and {Kalirai}, Jason and {Gaudi}, B. Scott and {Gilbert}, Karoline and {Girard}, Julien and {Grady}, Kevin and {Greene}, Jenny and {Guhathakurta}, Puragra and {Heinrich}, Chen and {Hemmati}, Shoubaneh and {Hendel}, David and {Henderson}, Calen and {Henning}, Thomas and {Hirata}, Christopher and {Ho}, Shirley and {Huff}, Eric and {Hutter}, Anne and {Jansen}, Rolf and {Jha}, Saurabh and {Johnson}, Samson and {Jones}, David and {Kasdin}, Jeremy and {Kelly}, Patrick and {Kirshner}, Robert and {Koekemoer}, Anton and {Kruk}, Jeffrey and {Lewis}, Nikole and {Macintosh}, Bruce and {Madau}, Piero and {Malhotra}, Sangeeta and {Mandel}, Kaisey and {Massara}, Elena and {Masters}, Daniel and {McEnery}, Julie and {McQuinn}, Kristen and {Melchior}, Peter and {Melton}, Mark and {Mennesson}, Bertrand and {Peeples}, Molly and {Penny}, Matthew and {Perlmutter}, Saul and {Pisani}, Alice and {Plazas}, Andr{\'e}s and {Poleski}, Radek and {Postman}, Marc and {Ranc}, Cl{\'e}ment and {Rauscher}, Bernard and {Rest}, Armin and {Roberge}, Aki and {Robertson}, Brant and {Rodney}, Steven and {Rhoads}, James and {Rhodes}, Jason and {Ryan}, Russell, Jr. and {Sahu}, Kailash and {Sand}, David and {Scolnic}, Dan and {Seth}, Anil and {Shvartzvald}, Yossi and {Siellez}, Karelle and {Smith}, Arfon and {Spergel}, David and {Stassun}, Keivan and {Street}, Rachel and {Strolger}, Louis-Gregory and {Szalay}, Alexander and {Trauger}, John and {Troxel}, M.~A. and {Turnbull}, Margaret and {van der Marel}, Roeland and {von der Linden}, Anja and {Wang}, Yun and {Weinberg}, David and {Williams}, Benjamin and {Windhorst}, Rogier and {Wollack}, Edward and {Wu}, Hao-Yi and {Yee}, Jennifer and {Zimmerman}, Neil},
        title = "{The Wide Field Infrared Survey Telescope: 100 Hubbles for the 2020s}",
      journal = {arXiv e-prints},
         year = 2019,
        month = feb,
          eid = {arXiv:1902.05569},
        pages = {arXiv:1902.05569},
          doi = {10.48550/arXiv.1902.05569},
archivePrefix = {arXiv},
       eprint = {1902.05569},
 primaryClass = {astro-ph.IM},
       adsurl = {https://ui.adsabs.harvard.edu/abs/2019arXiv190205569A}
}

@ARTICLE{LSST09,
       author = {{LSST Science Collaboration} and {Abell}, Paul A. and {Allison}, Julius and {Anderson}, Scott F. and {Andrew}, John R. and {Angel}, J. Roger P. and {Armus}, Lee and {Arnett}, David and {Asztalos}, S.~J. and {Axelrod}, Tim S. and {Bailey}, Stephen and {Ballantyne}, D.~R. and {Bankert}, Justin R. and {Barkhouse}, Wayne A. and {Barr}, Jeffrey D. and {Barrientos}, L. Felipe and {Barth}, Aaron J. and {Bartlett}, James G. and {Becker}, Andrew C. and {Becla}, Jacek and {Beers}, Timothy C. and {Bernstein}, Joseph P. and {Biswas}, Rahul and {Blanton}, Michael R. and {Bloom}, Joshua S. and {Bochanski}, John J. and {Boeshaar}, Pat and {Borne}, Kirk D. and {Bradac}, Marusa and {Brandt}, W.~N. and {Bridge}, Carrie R. and {Brown}, Michael E. and {Brunner}, Robert J. and {Bullock}, James S. and {Burgasser}, Adam J. and {Burge}, James H. and {Burke}, David L. and {Cargile}, Phillip A. and {Chandrasekharan}, Srinivasan and {Chartas}, George and {Chesley}, Steven R. and {Chu}, You-Hua and {Cinabro}, David and {Claire}, Mark W. and {Claver}, Charles F. and {Clowe}, Douglas and {Connolly}, A.~J. and {Cook}, Kem H. and {Cooke}, Jeff and {Cooray}, Asantha and {Covey}, Kevin R. and {Culliton}, Christopher S. and {de Jong}, Roelof and {de Vries}, Willem H. and {Debattista}, Victor P. and {Delgado}, Francisco and {Dell'Antonio}, Ian P. and {Dhital}, Saurav and {Di Stefano}, Rosanne and {Dickinson}, Mark and {Dilday}, Benjamin and {Djorgovski}, S.~G. and {Dobler}, Gregory and {Donalek}, Ciro and {Dubois-Felsmann}, Gregory and {Durech}, Josef and {Eliasdottir}, Ardis and {Eracleous}, Michael and {Eyer}, Laurent and {Falco}, Emilio E. and {Fan}, Xiaohui and {Fassnacht}, Christopher D. and {Ferguson}, Harry C. and {Fernandez}, Yanga R. and {Fields}, Brian D. and {Finkbeiner}, Douglas and {Figueroa}, Eduardo E. and {Fox}, Derek B. and {Francke}, Harold and {Frank}, James S. and {Frieman}, Josh and {Fromenteau}, Sebastien and {Furqan}, Muhammad and {Galaz}, Gaspar and {Gal-Yam}, A. and {Garnavich}, Peter and {Gawiser}, Eric and {Geary}, John and {Gee}, Perry and {Gibson}, Robert R. and {Gilmore}, Kirk and {Grace}, Emily A. and {Green}, Richard F. and {Gressler}, William J. and {Grillmair}, Carl J. and {Habib}, Salman and {Haggerty}, J.~S. and {Hamuy}, Mario and {Harris}, Alan W. and {Hawley}, Suzanne L. and {Heavens}, Alan F. and {Hebb}, Leslie and {Henry}, Todd J. and {Hileman}, Edward and {Hilton}, Eric J. and {Hoadley}, Keri and {Holberg}, J.~B. and {Holman}, Matt J. and {Howell}, Steve B. and {Infante}, Leopoldo and {Ivezic}, Zeljko and {Jacoby}, Suzanne H. and {Jain}, Bhuvnesh and {R} and {Jedicke} and {Jee}, M. James and {Garrett Jernigan}, J. and {Jha}, Saurabh W. and {Johnston}, Kathryn V. and {Jones}, R. Lynne and {Juric}, Mario and {Kaasalainen}, Mikko and {Styliani} and {Kafka} and {Kahn}, Steven M. and {Kaib}, Nathan A. and {Kalirai}, Jason and {Kantor}, Jeff and {Kasliwal}, Mansi M. and {Keeton}, Charles R. and {Kessler}, Richard and {Knezevic}, Zoran and {Kowalski}, Adam and {Krabbendam}, Victor L. and {Krughoff}, K. Simon and {Kulkarni}, Shrinivas and {Kuhlman}, Stephen and {Lacy}, Mark and {Lepine}, Sebastien and {Liang}, Ming and {Lien}, Amy and {Lira}, Paulina and {Long}, Knox S. and {Lorenz}, Suzanne and {Lotz}, Jennifer M. and {Lupton}, R.~H. and {Lutz}, Julie and {Macri}, Lucas M. and {Mahabal}, Ashish A. and {Mandelbaum}, Rachel and {Marshall}, Phil and {May}, Morgan and {McGehee}, Peregrine M. and {Meadows}, Brian T. and {Meert}, Alan and {Milani}, Andrea and {Miller}, Christopher J. and {Miller}, Michelle and {Mills}, David and {Minniti}, Dante and {Monet}, David and {Mukadam}, Anjum S. and {Nakar}, Ehud and {Neill}, Douglas R. and {Newman}, Jeffrey A. and {Nikolaev}, Sergei and {Nordby}, Martin and {O'Connor}, Paul and {Oguri}, Masamune and {Oliver}, John and {Olivier}, Scot S. and {Olsen}, Julia K. and {Olsen}, Knut and {Olszewski}, Edward W. and {Oluseyi}, Hakeem and {Padilla}, Nelson D. and {Parker}, Alex and {Pepper}, Joshua and {Peterson}, John R. and {Petry}, Catherine and {Pinto}, Philip A. and {Pizagno}, James L. and {Popescu}, Bogdan and {Prsa}, Andrej and {Radcka}, Veljko and {Raddick}, M. Jordan and {Rasmussen}, Andrew and {Rau}, Arne and {Rho}, Jeonghee and {Rhoads}, James E. and {Richards}, Gordon T. and {Ridgway}, Stephen T. and {Robertson}, Brant E. and {Roskar}, Rok and {Saha}, Abhijit and {Sarajedini}, Ata and {Scannapieco}, Evan and {Schalk}, Terry and {Schindler}, Rafe and {Schmidt}, Samuel and {Schmidt}, Sarah and {Schneider}, Donald P. and {Schumacher}, German and {Scranton}, Ryan and {Sebag}, Jacques and {Seppala}, Lynn G. and {Shemmer}, Ohad and {Simon}, Joshua D. and {Sivertz}, M. and {Smith}, Howard A. and {Allyn Smith}, J. and {Smith}, Nathan and {Spitz}, Anna H. and {Stanford}, Adam and {Stassun}, Keivan G. and {Strader}, Jay and {Strauss}, Michael A. and {Stubbs}, Christopher W. and {Sweeney}, Donald W. and {Szalay}, Alex and {Szkody}, Paula and {Takada}, Masahiro and {Thorman}, Paul and {Trilling}, David E. and {Trimble}, Virginia and {Tyson}, Anthony and {Van Berg}, Richard and {Vanden Berk}, Daniel and {VanderPlas}, Jake and {Verde}, Licia and {Vrsnak}, Bojan and {Walkowicz}, Lucianne M. and {Wandelt}, Benjamin D. and {Wang}, Sheng and {Wang}, Yun and {Warner}, Michael and {Wechsler}, Risa H. and {West}, Andrew A. and {Wiecha}, Oliver and {Williams}, Benjamin F. and {Willman}, Beth and {Wittman}, David and {Wolff}, Sidney C. and {Wood-Vasey}, W. Michael and {Wozniak}, Przemek and {Young}, Patrick and {Zentner}, Andrew and {Zhan}, Hu},
        title = "{LSST Science Book, Version 2.0}",
      journal = {arXiv e-prints},
         year = 2009,
        month = dec,
          eid = {arXiv:0912.0201},
        pages = {arXiv:0912.0201},
          doi = {10.48550/arXiv.0912.0201},
archivePrefix = {arXiv},
       eprint = {0912.0201},
 primaryClass = {astro-ph.IM},
       adsurl = {https://ui.adsabs.harvard.edu/abs/2009arXiv0912.0201L}
}

@ARTICLE{LSST18,
       author = {{The LSST Dark Energy Science Collaboration} and {Mandelbaum}, Rachel and {Eifler}, Tim and {Hlo{\v{z}}ek}, Ren{\'e}e and {Collett}, Thomas and {Gawiser}, Eric and {Scolnic}, Daniel and {Alonso}, David and {Awan}, Humna and {Biswas}, Rahul and {Blazek}, Jonathan and {Burchat}, Patricia and {Chisari}, Nora Elisa and {Dell'Antonio}, Ian and {Digel}, Seth and {Frieman}, Josh and {Goldstein}, Daniel A. and {Hook}, Isobel and {Ivezi{\'c}}, {\v{Z}}eljko and {Kahn}, Steven M. and {Kamath}, Sowmya and {Kirkby}, David and {Kitching}, Thomas and {Krause}, Elisabeth and {Leget}, Pierre-Fran{\c{c}}ois and {Marshall}, Philip J. and {Meyers}, Joshua and {Miyatake}, Hironao and {Newman}, Jeffrey A. and {Nichol}, Robert and {Rykoff}, Eli and {Sanchez}, F. Javier and {Slosar}, An{\v{z}}e and {Sullivan}, Mark and {Troxel}, M.~A.},
        title = "{The LSST Dark Energy Science Collaboration (DESC) Science Requirements Document}",
      journal = {arXiv e-prints},
         year = 2018,
        month = sep,
          eid = {arXiv:1809.01669},
        pages = {arXiv:1809.01669},
          doi = {10.48550/arXiv.1809.01669},
archivePrefix = {arXiv},
       eprint = {1809.01669},
 primaryClass = {astro-ph.CO},
       adsurl = {https://ui.adsabs.harvard.edu/abs/2018arXiv180901669T}
}

@ARTICLE{DESC12,
       author = {{LSST Dark Energy Science Collaboration}},
        title = "{Large Synoptic Survey Telescope: Dark Energy Science Collaboration}",
      journal = {arXiv e-prints},
         year = 2012,
        month = nov,
          eid = {arXiv:1211.0310},
        pages = {arXiv:1211.0310},
          doi = {10.48550/arXiv.1211.0310},
archivePrefix = {arXiv},
       eprint = {1211.0310},
 primaryClass = {astro-ph.CO},
       adsurl = {https://ui.adsabs.harvard.edu/abs/2012arXiv1211.0310L}
}

@ARTICLE{LSSTDC221,
       author = {{LSST Dark Energy Science Collaboration (LSST DESC)} and {Abolfathi}, Bela and {Alonso}, David and {Armstrong}, Robert and {Aubourg}, {\'E}ric and {Awan}, Humna and {Babuji}, Yadu N. and {Bauer}, Franz Erik and {Bean}, Rachel and {Beckett}, George and {Biswas}, Rahul and {Bogart}, Joanne R. and {Boutigny}, Dominique and {Chard}, Kyle and {Chiang}, James and {Claver}, Chuck F. and {Cohen-Tanugi}, Johann and {Combet}, C{\'e}line and {Connolly}, Andrew J. and {Daniel}, Scott F. and {Digel}, Seth W. and {Drlica-Wagner}, Alex and {Dubois}, Richard and {Gangler}, Emmanuel and {Gawiser}, Eric and {Glanzman}, Thomas and {Gris}, Phillipe and {Habib}, Salman and {Hearin}, Andrew P. and {Heitmann}, Katrin and {Hernandez}, Fabio and {Hlo{\v{z}}ek}, Ren{\'e}e and {Hollowed}, Joseph and {Ishak}, Mustapha and {Ivezi{\'c}}, {\v{Z}}eljko and {Jarvis}, Mike and {Jha}, Saurabh W. and {Kahn}, Steven M. and {Kalmbach}, J. Bryce and {Kelly}, Heather M. and {Kovacs}, Eve and {Korytov}, Danila and {Krughoff}, K. Simon and {Lage}, Craig S. and {Lanusse}, Fran{\c{c}}ois and {Larsen}, Patricia and {Le Guillou}, Laurent and {Li}, Nan and {Longley}, Emily Phillips and {Lupton}, Robert H. and {Mandelbaum}, Rachel and {Mao}, Yao-Yuan and {Marshall}, Phil and {Meyers}, Joshua E. and {Moniez}, Marc and {Morrison}, Christopher B. and {Nomerotski}, Andrei and {O'Connor}, Paul and {Park}, HyeYun and {Park}, Ji Won and {Peloton}, Julien and {Perrefort}, Daniel and {Perry}, James and {Plaszczynski}, St{\'e}phane and {Pope}, Adrian and {Rasmussen}, Andrew and {Reil}, Kevin and {Roodman}, Aaron J. and {Rykoff}, Eli S. and {S{\'a}nchez}, F. Javier and {Schmidt}, Samuel J. and {Scolnic}, Daniel and {Stubbs}, Christopher W. and {Tyson}, J. Anthony and {Uram}, Thomas D. and {Villarreal}, Antonia Sierra and {Walter}, Christopher W. and {Wiesner}, Matthew P. and {Wood-Vasey}, W. Michael and {Zuntz}, Joe},
        title = "{The LSST DESC DC2 Simulated Sky Survey}",
      journal = {\apjs},
         year = 2021,
        month = mar,
       volume = {253},
       number = {1},
          eid = {31},
        pages = {31},
          doi = {10.3847/1538-4365/abd62c},
archivePrefix = {arXiv},
       eprint = {2010.05926},
 primaryClass = {astro-ph.IM},
       adsurl = {https://ui.adsabs.harvard.edu/abs/2021ApJS..253...31L}
}

@ARTICLE{Heitmann19,
       author = {{Heitmann}, Katrin and {Finkel}, Hal and {Pope}, Adrian and {Morozov}, Vitali and {Frontiere}, Nicholas and {Habib}, Salman and {Rangel}, Esteban and {Uram}, Thomas and {Korytov}, Danila and {Child}, Hillary and {Flender}, Samuel and {Insley}, Joe and {Rizzi}, Silvio},
        title = "{The Outer Rim Simulation: A Path to Many-core Supercomputers}",
      journal = {\apjs},
         year = 2019,
        month = nov,
       volume = {245},
       number = {1},
          eid = {16},
        pages = {16},
          doi = {10.3847/1538-4365/ab4da1},
archivePrefix = {arXiv},
       eprint = {1904.11970},
 primaryClass = {astro-ph.CO},
       adsurl = {https://ui.adsabs.harvard.edu/abs/2019ApJS..245...16H}
}

@ARTICLE{Korytov19,
       author = {{Korytov}, Danila and {Hearin}, Andrew and {Kovacs}, Eve and {Larsen}, Patricia and {Rangel}, Esteban and {Hollowed}, Joseph and {Benson}, Andrew J. and {Heitmann}, Katrin and {Mao}, Yao-Yuan and {Bahmanyar}, Anita and {Chang}, Chihway and {Campbell}, Duncan and {DeRose}, Joseph and {Finkel}, Hal and {Frontiere}, Nicholas and {Gawiser}, Eric and {Habib}, Salman and {Joachimi}, Benjamin and {Lanusse}, Fran{\c{c}}ois and {Li}, Nan and {Mandelbaum}, Rachel and {Morrison}, Christopher and {Newman}, Jeffrey A. and {Pope}, Adrian and {Rykoff}, Eli and {Simet}, Melanie and {To}, Chun-Hao and {Vikraman}, Vinu and {Wechsler}, Risa H. and {White}, Martin and {(The LSST Dark Energy Science Collaboration}},
        title = "{CosmoDC2: A Synthetic Sky Catalog for Dark Energy Science with LSST}",
      journal = {\apjs},
         year = 2019,
        month = dec,
       volume = {245},
       number = {2},
          eid = {26},
        pages = {26},
          doi = {10.3847/1538-4365/ab510c},
archivePrefix = {arXiv},
       eprint = {1907.06530},
 primaryClass = {astro-ph.CO},
       adsurl = {https://ui.adsabs.harvard.edu/abs/2019ApJS..245...26K}
}

@ARTICLE{Slob24,
       author = {{Slob}, Martje and {Kriek}, Mariska and {Beverage}, Aliza G. and {Suess}, Katherine A. and {Barro}, Guillermo and {Bezanson}, Rachel and {Brammer}, Gabriel and {Cheng}, Chloe M. and {Conroy}, Charlie and {de Graaff}, Anna and {F{\"o}rster Schreiber}, Natascha M. and {Franx}, Marijn and {Lorenz}, Brian and {Mancera Pi{\~n}a}, Pavel E. and {Marchesini}, Danilo and {Muzzin}, Adam and {Newman}, Andrew B. and {Price}, Sedona H. and {Shapley}, Alice E. and {Stefanon}, Mauro and {van Dokkum}, Pieter and {Weisz}, Daniel R.},
        title = "{The JWST-SUSPENSE Ultradeep Spectroscopic Program: Survey Overview and Star-Formation Histories of Quiescent Galaxies at 1 < z < 3}",
      journal = {arXiv e-prints},
         year = 2024,
        month = apr,
          eid = {arXiv:2404.12432},
        pages = {arXiv:2404.12432},
          doi = {10.48550/arXiv.2404.12432},
archivePrefix = {arXiv},
       eprint = {2404.12432},
 primaryClass = {astro-ph.GA},
       adsurl = {https://ui.adsabs.harvard.edu/abs/2024arXiv240412432S}
}

@ARTICLE{Choi14,
       author = {{Choi}, Jieun and {Conroy}, Charlie and {Moustakas}, John and {Graves}, Genevieve J. and {Holden}, Bradford P. and {Brodwin}, Mark and {Brown}, Michael J.~I. and {van Dokkum}, Pieter G.},
        title = "{The Assembly Histories of Quiescent Galaxies since z = 0.7 from Absorption Line Spectroscopy}",
      journal = {\apj},
         year = 2014,
        month = sep,
       volume = {792},
       number = {2},
          eid = {95},
        pages = {95},
          doi = {10.1088/0004-637X/792/2/95},
archivePrefix = {arXiv},
       eprint = {1403.4932},
 primaryClass = {astro-ph.GA},
       adsurl = {https://ui.adsabs.harvard.edu/abs/2014ApJ...792...95C}
}

@ARTICLE{Beverage24,
       author = {{Beverage}, Aliza G. and {Slob}, Martje and {Kriek}, Mariska and {Conroy}, Charlie and {Barro}, Guillermo and {Bezanson}, Rachel and {Brammer}, Gabriel and {Cheng}, Chloe M. and {de Graaff}, Anna and {F{\"o}rster Schreiber}, Natascha M. and {Franx}, Marijn and {Lorenz}, Brian and {Mancera Pi{\~n}a}, Pavel E. and {Marchesini}, Danilo and {Muzzin}, Adam and {Newman}, Andrew B. and {Price}, Sedona H. and {Shapley}, Alice E. and {Stefanon}, Mauro and {Suess}, Katherine A. and {van Dokkum}, Pieter and {Weinberg}, David and {Weisz}, Daniel R.},
        title = "{Carbon and Iron Deficiencies in Quiescent Galaxies at z=1-3 from JWST-SUSPENSE: Implications for the Formation Histories of Massive Galaxies}",
      journal = {arXiv e-prints},
         year = 2024,
        month = jul,
          eid = {arXiv:2407.02556},
        pages = {arXiv:2407.02556},
          doi = {10.48550/arXiv.2407.02556},
archivePrefix = {arXiv},
       eprint = {2407.02556},
 primaryClass = {astro-ph.GA},
       adsurl = {https://ui.adsabs.harvard.edu/abs/2024arXiv240702556B}
}

@ARTICLE{Zhuang23,
       author = {{Zhuang}, Zhuyun and {Leethochawalit}, Nicha and {Kirby}, Evan N. and {Nightingale}, J.~W. and {Steidel}, Charles C. and {Glazebrook}, Karl and {Barone}, Tania M. and {Skobe}, Hannah and {Sweet}, Sarah M. and {Nanayakkara}, Themiya and {Allen}, Rebecca J. and {Vasan}, Keerthi G.~C. and {Jones}, Tucker and {Kacprzak}, Glenn G. and {Tran}, Kim-Vy H. and {Jacobs}, Colin},
        title = "{A Glimpse of the Stellar Populations and Elemental Abundances of Gravitationally Lensed, Quiescent Galaxies at z {\ensuremath{\gtrsim}} 1 with Keck Deep Spectroscopy}",
      journal = {\apj},
         year = 2023,
        month = may,
       volume = {948},
       number = {2},
          eid = {132},
        pages = {132},
          doi = {10.3847/1538-4357/acc79b},
archivePrefix = {arXiv},
       eprint = {2212.04731},
 primaryClass = {astro-ph.GA},
       adsurl = {https://ui.adsabs.harvard.edu/abs/2023ApJ...948..132Z}
}

@ARTICLE{Khullar22,
       author = {{Khullar}, Gourav and {Bayliss}, Matthew B. and {Gladders}, Michael D. and {Kim}, Keunho J. and {Calzadilla}, Michael S. and {Strazzullo}, Veronica and {Bleem}, Lindsey E. and {Mahler}, Guillaume and {McDonald}, Michael and {Floyd}, Benjamin and {Reichardt}, Christian L. and {Ruppin}, Florian and {Saro}, Alexandro and {Sharon}, Keren and {Somboonpanyakul}, Taweewat and {Stalder}, Brian and {Stark}, Antony A.},
        title = "{Synthesizing Stellar Populations in South Pole Telescope Galaxy Clusters. I. Ages of Quiescent Member Galaxies at 0.3 < z < 1.4}",
      journal = {\apj},
         year = 2022,
        month = aug,
       volume = {934},
       number = {2},
          eid = {177},
        pages = {177},
          doi = {10.3847/1538-4357/ac7c0c},
archivePrefix = {arXiv},
       eprint = {2111.09318},
 primaryClass = {astro-ph.GA},
       adsurl = {https://ui.adsabs.harvard.edu/abs/2022ApJ...934..177K}
}

@ARTICLE{Marsan22,
       author = {{Marsan}, Z. Cemile and {Muzzin}, Adam and {Marchesini}, Danilo and {Stefanon}, Mauro and {Martis}, Nicholas and {Annunziatella}, Marianna and {Chan}, Jeffrey C.~C. and {Cooper}, Michael C. and {Forrest}, Ben and {Gomez}, Percy and {McConachie}, Ian and {Wilson}, Gillian},
        title = "{The Number Densities and Stellar Populations of Massive Galaxies at 3 < z < 6: A Diverse, Rapidly Forming Population in the Early Universe}",
      journal = {\apj},
         year = 2022,
        month = jan,
       volume = {924},
       number = {1},
          eid = {25},
        pages = {25},
          doi = {10.3847/1538-4357/ac312a},
archivePrefix = {arXiv},
       eprint = {2010.04725},
 primaryClass = {astro-ph.GA},
       adsurl = {https://ui.adsabs.harvard.edu/abs/2022ApJ...924...25M}
}

@ARTICLE{Padmanabhan07,
       author = {{Padmanabhan}, Nikhil and {Schlegel}, David J. and {Seljak}, Uro{\v{s}} and {Makarov}, Alexey and {Bahcall}, Neta A. and {Blanton}, Michael R. and {Brinkmann}, Jonathan and {Eisenstein}, Daniel J. and {Finkbeiner}, Douglas P. and {Gunn}, James E. and {Hogg}, David W. and {Ivezi{\'c}}, {\v{Z}}eljko and {Knapp}, Gillian R. and {Loveday}, Jon and {Lupton}, Robert H. and {Nichol}, Robert C. and {Schneider}, Donald P. and {Strauss}, Michael A. and {Tegmark}, Max and {York}, Donald G.},
        title = "{The clustering of luminous red galaxies in the Sloan Digital Sky Survey imaging data}",
      journal = {\mnras},
         year = 2007,
        month = jul,
       volume = {378},
       number = {3},
        pages = {852-872},
          doi = {10.1111/j.1365-2966.2007.11593.x},
archivePrefix = {arXiv},
       eprint = {astro-ph/0605302},
 primaryClass = {astro-ph},
       adsurl = {https://ui.adsabs.harvard.edu/abs/2007MNRAS.378..852P}
}

@ARTICLE{Yuan24,
       author = {{Yuan}, Sihan and {Blake}, Chris and {Krolewski}, Alex and {Lange}, Johannes and {Elvin-Poole}, Jack and {Leauthaud}, Alexie and {DeRose}, Joseph and {Aguilar}, Jessica Nicole and {Ahlen}, Steven and {Beltz-Mohrmann}, Gillian and {Brooks}, David and {Claybaugh}, Todd and {de la Macorra}, Axel and {Doel}, Peter and {Emas}, Ni Putu Audita Placida and {Ferraro}, Simone and {Forero-Romero}, Jaime E. and {Garcia-Quintero}, Cristhian and {Gazta{\~n}aga}, Enrique and {Gontcho}, Satya Gontcho A. and {Hadzhiyska}, Boryana and {Heydenreich}, Sven and {Honscheid}, Klaus and {Ishak}, Mustapha and {Joudaki}, Shahab and {Jullo}, Eric and {Kisner}, Theodore and {Kremin}, Anthony and {Lambert}, Andrew and {Landriau}, Martin and {Manera}, Marc and {Meisner}, Aaron and {Miquel}, Ramon and {Nie}, Jundan and {Palanque-Delabrouille}, Nathalie and {Poppett}, Claire and {Porredon}, Anna and {Rezaie}, Mehdi and {Ross}, Ashley J. and {Rossi}, Graziano and {Ruggeri}, Rossana and {Sanchez}, Eusebio and {Saulder}, Christoph and {Seo}, Hee-Jong and {Silber}, Joseph Harry and {Tarl{\'n}}, Gregory and {Vargas-Maga{\~n}a}, Mariana and {Weaver}, Benjamin Alan and {Xhakaj}, Enia and {Zhou}, Zhimin and {Zou}, Hu},
        title = "{Redshift evolution and covariances for joint lensing and clustering studies with DESI Y1}",
      journal = {\mnras},
         year = 2024,
        month = sep,
       volume = {533},
       number = {1},
        pages = {589-607},
          doi = {10.1093/mnras/stae1792},
archivePrefix = {arXiv},
       eprint = {2403.00915},
 primaryClass = {astro-ph.CO},
       adsurl = {https://ui.adsabs.harvard.edu/abs/2024MNRAS.533..589Y}
}

@ARTICLE{Sailer24,
       author = {{Sailer}, Noah and {Kim}, Joshua and {Ferraro}, Simone and {Madhavacheril}, Mathew S. and {White}, Martin and {Abril-Cabezas}, Irene and {Aguilar}, Jessica Nicole and {Ahlen}, Steven and {Bond}, J. Richard and {Brooks}, David and {Burtin}, Etienne and {Calabrese}, Erminia and {Chen}, Shi-Fan and {Choi}, Steve K. and {Claybaugh}, Todd and {Dawson}, Kyle and {de la Macorra}, Axel and {DeRose}, Joseph and {Dey}, Arjun and {Dey}, Biprateep and {Doel}, Peter and {Dunkley}, Jo and {Embil-Villagra}, Carmen and {Farren}, Gerrit S. and {Font-Ribera}, Andreu and {Forero-Romero}, Jaime E. and {Gazta{\~n}aga}, Enrique and {Gluscevic}, Vera and {Gontcho}, Satya Gontcho A and {Honscheid}, Klaus and {Howlett}, Cullan and {Juneau}, Stephanie and {Kirkby}, David and {Kisner}, Theodore and {Kremin}, Anthony and {Landriau}, Martin and {Le Guillou}, Laurent and {Levi}, Michael and {Manera}, Marc and {Meisner}, Aaron and {Miquel}, Ramon and {Moodley}, Kavilan and {Moustakas}, John and {Niemack}, Michael D. and {Niz}, Gustavo and {Palanque-Delabrouille}, Nathalie and {Percival}, Will and {Prada}, Francisco and {Qu}, Frank J. and {Rossi}, Graziano and {Sanchez}, Eusebio and {Schaan}, Emmanuel and {Schlafly}, Edward and {Schlegel}, David and {Schubnell}, Michael and {Sehgal}, Neelima and {Seo}, Hee-Jong and {Sherwin}, Blake and {Sif{\'o}n}, Crist{\'o}bal and {Sprayberry}, David and {Staggs}, Suzanne T. and {Tarl{\'e}}, Gregory and {Weaver}, Benjamin Alan and {Y{\`e}che}, Christophe and {Zhou}, Rongpu and {Zou}, Hu},
        title = "{Cosmological constraints from the cross-correlation of DESI Luminous Red Galaxies with CMB lensing from Planck PR4 and ACT DR6}",
      journal = {arXiv e-prints},
         year = 2024,
        month = jul,
          eid = {arXiv:2407.04607},
        pages = {arXiv:2407.04607},
          doi = {10.48550/arXiv.2407.04607},
archivePrefix = {arXiv},
       eprint = {2407.04607},
 primaryClass = {astro-ph.CO},
       adsurl = {https://ui.adsabs.harvard.edu/abs/2024arXiv240704607S}
}

@ARTICLE{White22,
       author = {{White}, Martin and {Zhou}, Rongpu and {DeRose}, Joseph and {Ferraro}, Simone and {Chen}, Shi-Fan and {Kokron}, Nickolas and {Bailey}, Stephen and {Brooks}, David and {Garc{\'\i}a-Bellido}, Juan and {Guy}, Julien and {Honscheid}, Klaus and {Kehoe}, Robert and {Kremin}, Anthony and {Levi}, Michael and {Palanque-Delabrouille}, Nathalie and {Poppett}, Claire and {Schlegel}, David and {Tarle}, Gregory},
        title = "{Cosmological constraints from the tomographic cross-correlation of DESI Luminous Red Galaxies and Planck CMB lensing}",
      journal = {\jcap},
         year = 2022,
        month = feb,
       volume = {2022},
       number = {2},
          eid = {007},
        pages = {007},
          doi = {10.1088/1475-7516/2022/02/007},
archivePrefix = {arXiv},
       eprint = {2111.09898},
 primaryClass = {astro-ph.CO},
       adsurl = {https://ui.adsabs.harvard.edu/abs/2022JCAP...02..007W}
}

@ARTICLE{Percival07,
       author = {{Percival}, Will J. and {Cole}, Shaun and {Eisenstein}, Daniel J. and {Nichol}, Robert C. and {Peacock}, John A. and {Pope}, Adrian C. and {Szalay}, Alexander S.},
        title = "{Measuring the Baryon Acoustic Oscillation scale using the Sloan Digital Sky Survey and 2dF Galaxy Redshift Survey}",
      journal = {\mnras},
         year = 2007,
        month = nov,
       volume = {381},
       number = {3},
        pages = {1053-1066},
          doi = {10.1111/j.1365-2966.2007.12268.x},
archivePrefix = {arXiv},
       eprint = {0705.3323},
 primaryClass = {astro-ph},
       adsurl = {https://ui.adsabs.harvard.edu/abs/2007MNRAS.381.1053P}
}

@ARTICLE{Rosell22,
       author = {{Carnero Rosell}, A. and {Rodriguez-Monroy}, M. and {Crocce}, M. and {Elvin-Poole}, J. and {Porredon}, A. and {Ferrero}, I. and {Mena-Fern{\'a}ndez}, J. and {Cawthon}, R. and {De Vicente}, J. and {Gaztanaga}, E. and {Ross}, A.~J. and {Sanchez}, E. and {Sevilla-Noarbe}, I. and {Alves}, O. and {Andrade-Oliveira}, F. and {Asorey}, J. and {Avila}, S. and {Brandao-Souza}, A. and {Camacho}, H. and {Chan}, K.~C. and {Fert{\'e}}, A. and {Muir}, J. and {Riquelme}, W. and {Rosenfeld}, R. and {Sanchez Cid}, D. and {Hartley}, W.~G. and {Weaverdyck}, N. and {Abbott}, T. and {Aguena}, M. and {Allam}, S. and {Annis}, J. and {Bertin}, E. and {Brooks}, D. and {Buckley-Geer}, E. and {Burke}, D. and {Calcino}, J. and {Carollo}, D. and {Carrasco Kind}, M. and {Carretero}, J. and {Castander}, F. and {Choi}, A. and {Conselice}, C. and {Costanzi}, M. and {da Costa}, L. and {da Silva Pereira}, M.~E. and {Davis}, T. and {Desai}, S. and {Diehl}, H.~T. and {Doel}, P. and {Drlica-Wagner}, A. and {Eckert}, K. and {Everett}, S. and {Evrard}, A. and {Flaugher}, B. and {Fosalba}, P. and {Frieman}, J. and {Garcia-Bellido}, J. and {Gerdes}, D. and {Giannantonio}, T. and {Glazebrook}, K. and {Gruen}, D. and {Gruendl}, R. and {Gschwend}, J. and {Gutierrez}, G. and {Hinton}, S. and {Hollowood}, D. and {Honscheid}, K. and {Hoyle}, B. and {Huterer}, D. and {James}, D. and {Kim}, A. and {Krause}, E. and {Kuehn}, K. and {Lahav}, O. and {Lewis}, G. and {Lidman}, C. and {Lima}, M. and {Maia}, M. and {Malik}, U. and {Marshall}, J. and {Menanteau}, F. and {Miquel}, R. and {Mohr}, J. and {Moller}, A. and {Morgan}, R. and {Ogando}, R. and {Palmese}, A. and {Paz-Chinchon}, F. and {Percival}, W. and {Pieres}, A. and {Malag{\'o}n}, A. Plazas and {Roodman}, A. and {Scarpine}, V. and {Schubnell}, M. and {Serrano}, S. and {Sharp}, R. and {Sheldon}, E. and {Smith}, M. and {Soares-Santos}, M. and {Suchyta}, E. and {Swanson}, M. and {Tarle}, G. and {Thomas}, D. and {To}, C. and {Tucker}, B. and {Tucker}, D. and {Uddin}, S. and {Varga}, T.~N. and {Varga}, T.~N. and {DES Collaboration}},
        title = "{Dark Energy Survey Year 3 results: galaxy sample for BAO measurement}",
      journal = {\mnras},
         year = 2022,
        month = jan,
       volume = {509},
       number = {1},
        pages = {778-799},
          doi = {10.1093/mnras/stab2995},
archivePrefix = {arXiv},
       eprint = {2107.05477},
 primaryClass = {astro-ph.CO},
       adsurl = {https://ui.adsabs.harvard.edu/abs/2022MNRAS.509..778C}
}

@article{Eifler21,
    author = {Eifler, Tim and Miyatake, Hironao and Krause, Elisabeth and Heinrich, Chen and Miranda, Vivian and Hirata, Christopher and Xu, Jiachuan and Hemmati, Shoubaneh and Simet, Melanie and Capak, Peter and Choi, Ami and Doré, Olivier and Doux, Cyrille and Fang, Xiao and Hounsell, Rebekah and Huff, Eric and Huang, Hung-Jin and Jarvis, Mike and Kruk, Jeffrey and Masters, Dan and Rozo, Eduardo and Scolnic, Dan and Spergel, David N and Troxel, Michael and von der Linden, Anja and Wang, Yun and Weinberg, David H and Wenzl, Lukas and Wu, Hao-Yi},
    title = "{Cosmology with the Roman Space Telescope – multiprobe strategies}",
    journal = {Monthly Notices of the Royal Astronomical Society},
    volume = {507},
    number = {2},
    pages = {1746-1761},
    year = {2021},
    month = {07},
    issn = {0035-8711},
    doi = {10.1093/mnras/stab1762},
    url = {https://doi.org/10.1093/mnras/stab1762},
    eprint = {https://academic.oup.com/mnras/article-pdf/507/2/1746/40048399/stab1762.pdf},
}

@ARTICLE{Zhou23a,
       author = {{Zhou}, Rongpu and {Dey}, Biprateep and {Newman}, Jeffrey A. and {Eisenstein}, Daniel J. and {Dawson}, K. and {Bailey}, S. and {Berti}, A. and {Guy}, J. and {Lan}, Ting-Wen and {Zou}, H. and {Aguilar}, J. and {Ahlen}, S. and {Alam}, Shadab and {Brooks}, D. and {de la Macorra}, A. and {Dey}, A. and {Dhungana}, G. and {Fanning}, K. and {Font-Ribera}, A. and {Gontcho}, S. Gontcho A. and {Honscheid}, K. and {Ishak}, Mustapha and {Kisner}, T. and {Kov{\'a}cs}, A. and {Kremin}, A. and {Landriau}, M. and {Levi}, Michael E. and {Magneville}, C. and {Manera}, Marc and {Martini}, P. and {Meisner}, Aaron M. and {Miquel}, R. and {Moustakas}, J. and {Myers}, Adam D. and {Nie}, Jundan and {Palanque-Delabrouille}, N. and {Percival}, W.~J. and {Poppett}, C. and {Prada}, F. and {Raichoor}, A. and {Ross}, A.~J. and {Schlafly}, E. and {Schlegel}, D. and {Schubnell}, M. and {Tarl{\'e}}, Gregory and {Weaver}, B.~A. and {Wechsler}, R.~H. and {Y{\'e}che}, Christophe and {Zhou}, Zhimin},
        title = "{Target Selection and Validation of DESI Luminous Red Galaxies}",
      journal = {\aj},
         year = 2023,
        month = feb,
       volume = {165},
       number = {2},
          eid = {58},
        pages = {58},
          doi = {10.3847/1538-3881/aca5fb},
archivePrefix = {arXiv},
       eprint = {2208.08515},
 primaryClass = {astro-ph.CO},
       adsurl = {https://ui.adsabs.harvard.edu/abs/2023AJ....165...58Z}
}

@ARTICLE{Zhou23b,
       author = {{Zhou}, Rongpu and {Ferraro}, Simone and {White}, Martin and {DeRose}, Joseph and {Sailer}, Noah and {Aguilar}, Jessica and {Ahlen}, Steven and {Bailey}, Stephen and {Brooks}, David and {Claybaugh}, Todd and {Dawson}, Kyle and {de la Macorra}, Axel and {Dey}, Biprateep and {Doel}, Peter and {Font-Ribera}, Andreu and {Forero-Romero}, Jaime E. and {Gontcho A Gontcho}, Satya and {Guy}, Julien and {Kremin}, Anthony and {Lambert}, Andrew and {Le Guillou}, Laurent and {Levi}, Michael and {Magneville}, Christophe and {Manera}, Marc and {Meisner}, Aaron and {Miquel}, Ramon and {Moustakas}, John and {Myers}, Adam D. and {Newman}, Jeffrey A. and {Nie}, Jundan and {Percival}, Will and {Rezaie}, Mehdi and {Rossi}, Graziano and {Sanchez}, Eusebio and {Schlegel}, David and {Schubnell}, Michael and {Seo}, Hee-Jong and {Tarl{\'e}}, Gregory and {Zhou}, Zhimin},
        title = "{DESI luminous red galaxy samples for cross-correlations}",
      journal = {\jcap},
         year = 2023,
        month = nov,
       volume = {2023},
       number = {11},
          eid = {097},
        pages = {097},
          doi = {10.1088/1475-7516/2023/11/097},
archivePrefix = {arXiv},
       eprint = {2309.06443},
 primaryClass = {astro-ph.CO},
       adsurl = {https://ui.adsabs.harvard.edu/abs/2023JCAP...11..097Z}
}

@ARTICLE{Rykoff16,
       author = {{Rykoff}, E.~S. and {Rozo}, E. and {Hollowood}, D. and {Bermeo-Hernandez}, A. and {Jeltema}, T. and {Mayers}, J. and {Romer}, A.~K. and {Rooney}, P. and {Saro}, A. and {Vergara Cervantes}, C. and {Wechsler}, R.~H. and {Wilcox}, H. and {Abbott}, T.~M.~C. and {Abdalla}, F.~B. and {Allam}, S. and {Annis}, J. and {Benoit-L{\'e}vy}, A. and {Bernstein}, G.~M. and {Bertin}, E. and {Brooks}, D. and {Burke}, D.~L. and {Capozzi}, D. and {Carnero Rosell}, A. and {Carrasco Kind}, M. and {Castander}, F.~J. and {Childress}, M. and {Collins}, C.~A. and {Cunha}, C.~E. and {D'Andrea}, C.~B. and {da Costa}, L.~N. and {Davis}, T.~M. and {Desai}, S. and {Diehl}, H.~T. and {Dietrich}, J.~P. and {Doel}, P. and {Evrard}, A.~E. and {Finley}, D.~A. and {Flaugher}, B. and {Fosalba}, P. and {Frieman}, J. and {Glazebrook}, K. and {Goldstein}, D.~A. and {Gruen}, D. and {Gruendl}, R.~A. and {Gutierrez}, G. and {Hilton}, M. and {Honscheid}, K. and {Hoyle}, B. and {James}, D.~J. and {Kay}, S.~T. and {Kuehn}, K. and {Kuropatkin}, N. and {Lahav}, O. and {Lewis}, G.~F. and {Lidman}, C. and {Lima}, M. and {Maia}, M.~A.~G. and {Mann}, R.~G. and {Marshall}, J.~L. and {Martini}, P. and {Melchior}, P. and {Miller}, C.~J. and {Miquel}, R. and {Mohr}, J.~J. and {Nichol}, R.~C. and {Nord}, B. and {Ogando}, R. and {Plazas}, A.~A. and {Reil}, K. and {Sahl{\'e}n}, M. and {Sanchez}, E. and {Santiago}, B. and {Scarpine}, V. and {Schubnell}, M. and {Sevilla-Noarbe}, I. and {Smith}, R.~C. and {Soares-Santos}, M. and {Sobreira}, F. and {Stott}, J.~P. and {Suchyta}, E. and {Swanson}, M.~E.~C. and {Tarle}, G. and {Thomas}, D. and {Tucker}, D. and {Uddin}, S. and {Viana}, P.~T.~P. and {Vikram}, V. and {Walker}, A.~R. and {Zhang}, Y. and {DES Collaboration}},
        title = "{The RedMaPPer Galaxy Cluster Catalog From DES Science Verification Data}",
      journal = {\apjs},
         year = 2016,
        month = may,
       volume = {224},
       number = {1},
          eid = {1},
        pages = {1},
          doi = {10.3847/0067-0049/224/1/1},
archivePrefix = {arXiv},
       eprint = {1601.00621},
 primaryClass = {astro-ph.CO},
       adsurl = {https://ui.adsabs.harvard.edu/abs/2016ApJS..224....1R}
}

@ARTICLE{Oguri14,
       author = {{Oguri}, Masamune},
        title = "{A cluster finding algorithm based on the multiband identification of red sequence galaxies}",
      journal = {\mnras},
         year = 2014,
        month = oct,
       volume = {444},
       number = {1},
        pages = {147-161},
          doi = {10.1093/mnras/stu1446},
archivePrefix = {arXiv},
       eprint = {1407.4693},
 primaryClass = {astro-ph.CO},
       adsurl = {https://ui.adsabs.harvard.edu/abs/2014MNRAS.444..147O}
}

@article{Oguri17,
    author = {Oguri, Masamune and Lin, Yen-Ting and Lin, Sheng-Chieh and Nishizawa, Atsushi J and More, Anupreeta and More, Surhud and Hsieh, Bau-Ching and Medezinski, Elinor and Miyatake, Hironao and Jian, Hung-Yu and Lin, Lihwai and Takada, Masahiro and Okabe, Nobuhiro and Speagle, Joshua S and Coupon, Jean and Leauthaud, Alexie and Lupton, Robert H and Miyazaki, Satoshi and Price, Paul A and Tanaka, Masayuki and Chiu, I-Non and Komiyama, Yutaka and Okura, Yuki and Tanaka, Manobu M and Usuda, Tomonori},
    title = "{An optically-selected cluster catalog at redshift 0.1 \&lt; z \&lt; 1.1 from the Hyper Suprime-Cam Subaru Strategic Program S16A data}",
    journal = {Publications of the Astronomical Society of Japan},
    volume = {70},
    number = {SP1},
    pages = {S20},
    year = {2017},
    month = {06},
    issn = {0004-6264},
    doi = {10.1093/pasj/psx042},
    url = {https://doi.org/10.1093/pasj/psx042},
    eprint = {https://academic.oup.com/pasj/article-pdf/70/SP1/S20/54675166/pasj\_70\_sp1\_s20.pdf},
}

@ARTICLE{Aihara18,
       author = {{Aihara}, Hiroaki and {Arimoto}, Nobuo and {Armstrong}, Robert and {Arnouts}, St{\'e}phane and {Bahcall}, Neta A. and {Bickerton}, Steven and {Bosch}, James and {Bundy}, Kevin and {Capak}, Peter L. and {Chan}, James H.~H. and {Chiba}, Masashi and {Coupon}, Jean and {Egami}, Eiichi and {Enoki}, Motohiro and {Finet}, Francois and {Fujimori}, Hiroki and {Fujimoto}, Seiji and {Furusawa}, Hisanori and {Furusawa}, Junko and {Goto}, Tomotsugu and {Goulding}, Andy and {Greco}, Johnny P. and {Greene}, Jenny E. and {Gunn}, James E. and {Hamana}, Takashi and {Harikane}, Yuichi and {Hashimoto}, Yasuhiro and {Hattori}, Takashi and {Hayashi}, Masao and {Hayashi}, Yusuke and {He{\l}miniak}, Krzysztof G. and {Higuchi}, Ryo and {Hikage}, Chiaki and {Ho}, Paul T.~P. and {Hsieh}, Bau-Ching and {Huang}, Kuiyun and {Huang}, Song and {Ikeda}, Hiroyuki and {Imanishi}, Masatoshi and {Inoue}, Akio K. and {Iwasawa}, Kazushi and {Iwata}, Ikuru and {Jaelani}, Anton T. and {Jian}, Hung-Yu and {Kamata}, Yukiko and {Karoji}, Hiroshi and {Kashikawa}, Nobunari and {Katayama}, Nobuhiko and {Kawanomoto}, Satoshi and {Kayo}, Issha and {Koda}, Jin and {Koike}, Michitaro and {Kojima}, Takashi and {Komiyama}, Yutaka and {Konno}, Akira and {Koshida}, Shintaro and {Koyama}, Yusei and {Kusakabe}, Haruka and {Leauthaud}, Alexie and {Lee}, Chien-Hsiu and {Lin}, Lihwai and {Lin}, Yen-Ting and {Lupton}, Robert H. and {Mandelbaum}, Rachel and {Matsuoka}, Yoshiki and {Medezinski}, Elinor and {Mineo}, Sogo and {Miyama}, Shoken and {Miyatake}, Hironao and {Miyazaki}, Satoshi and {Momose}, Rieko and {More}, Anupreeta and {More}, Surhud and {Moritani}, Yuki and {Moriya}, Takashi J. and {Morokuma}, Tomoki and {Mukae}, Shiro and {Murata}, Ryoma and {Murayama}, Hitoshi and {Nagao}, Tohru and {Nakata}, Fumiaki and {Niida}, Mana and {Niikura}, Hiroko and {Nishizawa}, Atsushi J. and {Obuchi}, Yoshiyuki and {Oguri}, Masamune and {Oishi}, Yukie and {Okabe}, Nobuhiro and {Okamoto}, Sakurako and {Okura}, Yuki and {Ono}, Yoshiaki and {Onodera}, Masato and {Onoue}, Masafusa and {Osato}, Ken and {Ouchi}, Masami and {Price}, Paul A. and {Pyo}, Tae-Soo and {Sako}, Masao and {Sawicki}, Marcin and {Shibuya}, Takatoshi and {Shimasaku}, Kazuhiro and {Shimono}, Atsushi and {Shirasaki}, Masato and {Silverman}, John D. and {Simet}, Melanie and {Speagle}, Joshua and {Spergel}, David N. and {Strauss}, Michael A. and {Sugahara}, Yuma and {Sugiyama}, Naoshi and {Suto}, Yasushi and {Suyu}, Sherry H. and {Suzuki}, Nao and {Tait}, Philip J. and {Takada}, Masahiro and {Takata}, Tadafumi and {Tamura}, Naoyuki and {Tanaka}, Manobu M. and {Tanaka}, Masaomi and {Tanaka}, Masayuki and {Tanaka}, Yoko and {Terai}, Tsuyoshi and {Terashima}, Yuichi and {Toba}, Yoshiki and {Tominaga}, Nozomu and {Toshikawa}, Jun and {Turner}, Edwin L. and {Uchida}, Tomohisa and {Uchiyama}, Hisakazu and {Umetsu}, Keiichi and {Uraguchi}, Fumihiro and {Urata}, Yuji and {Usuda}, Tomonori and {Utsumi}, Yousuke and {Wang}, Shiang-Yu and {Wang}, Wei-Hao and {Wong}, Kenneth C. and {Yabe}, Kiyoto and {Yamada}, Yoshihiko and {Yamanoi}, Hitomi and {Yasuda}, Naoki and {Yeh}, Sherry and {Yonehara}, Atsunori and {Yuma}, Suraphong},
        title = "{The Hyper Suprime-Cam SSP Survey: Overview and survey design}",
      journal = {\pasj},
         year = 2018,
        month = jan,
       volume = {70},
          eid = {S4},
        pages = {S4},
          doi = {10.1093/pasj/psx066},
archivePrefix = {arXiv},
       eprint = {1704.05858},
 primaryClass = {astro-ph.IM},
       adsurl = {https://ui.adsabs.harvard.edu/abs/2018PASJ...70S...4A}
}

@ARTICLE{Vakili23,
       author = {{Vakili}, Mohammadjavad and {Hoekstra}, Henk and {Bilicki}, Maciej and {Fortuna}, Maria Cristina and {Kuijken}, Konrad and {Wright}, Angus H. and {Asgari}, Marika and {Brown}, Michael and {Dombrovskij}, Elisabeth and {Erben}, Thomas and {Giblin}, Benjamin and {Heymans}, Catherine and {Hildebrandt}, Hendrik and {Johnston}, Harry and {Joudaki}, Shahab and {Kannawadi}, Arun},
        title = "{Clustering of red sequence galaxies in the fourth data release of the Kilo-Degree Survey}",
      journal = {\aap},
         year = 2023,
        month = jul,
       volume = {675},
          eid = {A202},
        pages = {A202},
          doi = {10.1051/0004-6361/202039293},
archivePrefix = {arXiv},
       eprint = {2008.13154},
 primaryClass = {astro-ph.CO},
       adsurl = {https://ui.adsabs.harvard.edu/abs/2023A&A...675A.202V}
}

@ARTICLE{DES16,
       author = {{Dark Energy Survey Collaboration} and {Abbott}, T. and {Abdalla}, F.~B. and {Aleksi{\'c}}, J. and {Allam}, S. and {Amara}, A. and {Bacon}, D. and {Balbinot}, E. and {Banerji}, M. and {Bechtol}, K. and {Benoit-L{\'e}vy}, A. and {Bernstein}, G.~M. and {Bertin}, E. and {Blazek}, J. and {Bonnett}, C. and {Bridle}, S. and {Brooks}, D. and {Brunner}, R.~J. and {Buckley-Geer}, E. and {Burke}, D.~L. and {Caminha}, G.~B. and {Capozzi}, D. and {Carlsen}, J. and {Carnero-Rosell}, A. and {Carollo}, M. and {Carrasco-Kind}, M. and {Carretero}, J. and {Castander}, F.~J. and {Clerkin}, L. and {Collett}, T. and {Conselice}, C. and {Crocce}, M. and {Cunha}, C.~E. and {D'Andrea}, C.~B. and {da Costa}, L.~N. and {Davis}, T.~M. and {Desai}, S. and {Diehl}, H.~T. and {Dietrich}, J.~P. and {Dodelson}, S. and {Doel}, P. and {Drlica-Wagner}, A. and {Estrada}, J. and {Etherington}, J. and {Evrard}, A.~E. and {Fabbri}, J. and {Finley}, D.~A. and {Flaugher}, B. and {Foley}, R.~J. and {Fosalba}, P. and {Frieman}, J. and {Garc{\'\i}a-Bellido}, J. and {Gaztanaga}, E. and {Gerdes}, D.~W. and {Giannantonio}, T. and {Goldstein}, D.~A. and {Gruen}, D. and {Gruendl}, R.~A. and {Guarnieri}, P. and {Gutierrez}, G. and {Hartley}, W. and {Honscheid}, K. and {Jain}, B. and {James}, D.~J. and {Jeltema}, T. and {Jouvel}, S. and {Kessler}, R. and {King}, A. and {Kirk}, D. and {Kron}, R. and {Kuehn}, K. and {Kuropatkin}, N. and {Lahav}, O. and {Li}, T.~S. and {Lima}, M. and {Lin}, H. and {Maia}, M.~A.~G. and {Makler}, M. and {Manera}, M. and {Maraston}, C. and {Marshall}, J.~L. and {Martini}, P. and {McMahon}, R.~G. and {Melchior}, P. and {Merson}, A. and {Miller}, C.~J. and {Miquel}, R. and {Mohr}, J.~J. and {Morice-Atkinson}, X. and {Naidoo}, K. and {Neilsen}, E. and {Nichol}, R.~C. and {Nord}, B. and {Ogando}, R. and {Ostrovski}, F. and {Palmese}, A. and {Papadopoulos}, A. and {Peiris}, H.~V. and {Peoples}, J. and {Percival}, W.~J. and {Plazas}, A.~A. and {Reed}, S.~L. and {Refregier}, A. and {Romer}, A.~K. and {Roodman}, A. and {Ross}, A. and {Rozo}, E. and {Rykoff}, E.~S. and {Sadeh}, I. and {Sako}, M. and {S{\'a}nchez}, C. and {Sanchez}, E. and {Santiago}, B. and {Scarpine}, V. and {Schubnell}, M. and {Sevilla-Noarbe}, I. and {Sheldon}, E. and {Smith}, M. and {Smith}, R.~C. and {Soares-Santos}, M. and {Sobreira}, F. and {Soumagnac}, M. and {Suchyta}, E. and {Sullivan}, M. and {Swanson}, M. and {Tarle}, G. and {Thaler}, J. and {Thomas}, D. and {Thomas}, R.~C. and {Tucker}, D. and {Vieira}, J.~D. and {Vikram}, V. and {Walker}, A.~R. and {Wechsler}, R.~H. and {Weller}, J. and {Wester}, W. and {Whiteway}, L. and {Wilcox}, H. and {Yanny}, B. and {Zhang}, Y. and {Zuntz}, J.},
        title = "{The Dark Energy Survey: more than dark energy - an overview}",
      journal = {\mnras},
         year = 2016,
        month = aug,
       volume = {460},
       number = {2},
        pages = {1270-1299},
          doi = {10.1093/mnras/stw641},
archivePrefix = {arXiv},
       eprint = {1601.00329},
 primaryClass = {astro-ph.CO},
       adsurl = {https://ui.adsabs.harvard.edu/abs/2016MNRAS.460.1270D}
}

@ARTICLE{deJong13,
       author = {{de Jong}, Jelte T.~A. and {Verdoes Kleijn}, Gijs A. and {Kuijken}, Konrad H. and {Valentijn}, Edwin A.},
        title = "{The Kilo-Degree Survey}",
      journal = {Experimental Astronomy},
         year = 2013,
        month = jan,
       volume = {35},
       number = {1-2},
        pages = {25-44},
          doi = {10.1007/s10686-012-9306-1},
archivePrefix = {arXiv},
       eprint = {1206.1254},
 primaryClass = {astro-ph.CO},
       adsurl = {https://ui.adsabs.harvard.edu/abs/2013ExA....35...25D}
}

@ARTICLE{SExtractor,
       author = {{Bertin}, E. and {Arnouts}, S.},
        title = "{SExtractor: Software for source extraction.}",
      journal = {\aaps},
         year = 1996,
        month = jun,
       volume = {117},
        pages = {393-404},
          doi = {10.1051/aas:1996164},
       adsurl = {https://ui.adsabs.harvard.edu/abs/1996A&AS..117..393B}
}

\appendix 
\section{IRMaGiC Photometric Redshift Performance on Truth Catalogs}\label{appendix}

\begin{figure*}[h]
    \centering
    \includegraphics[width=0.48\linewidth]{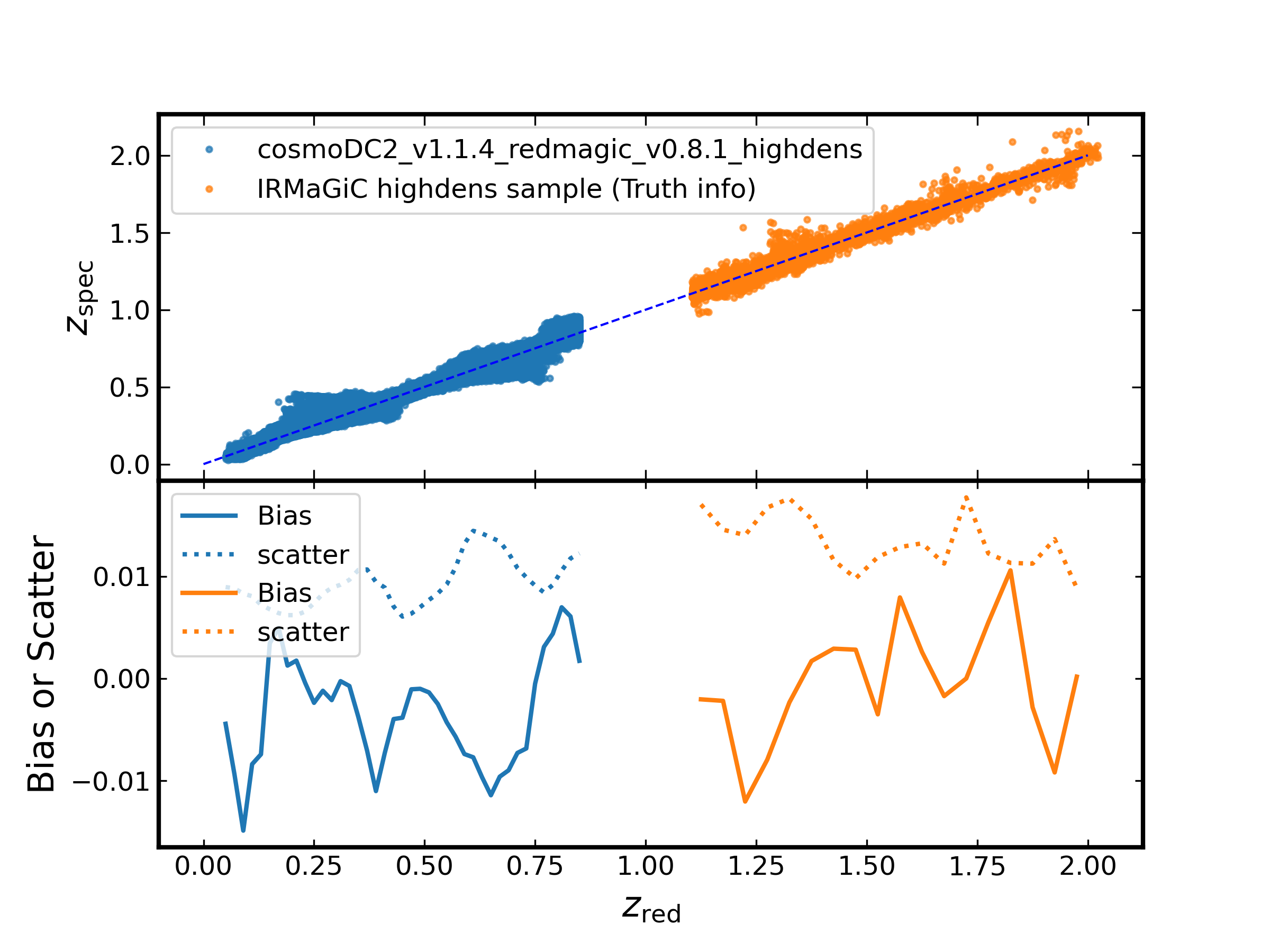}
    \includegraphics[width=0.48\linewidth]{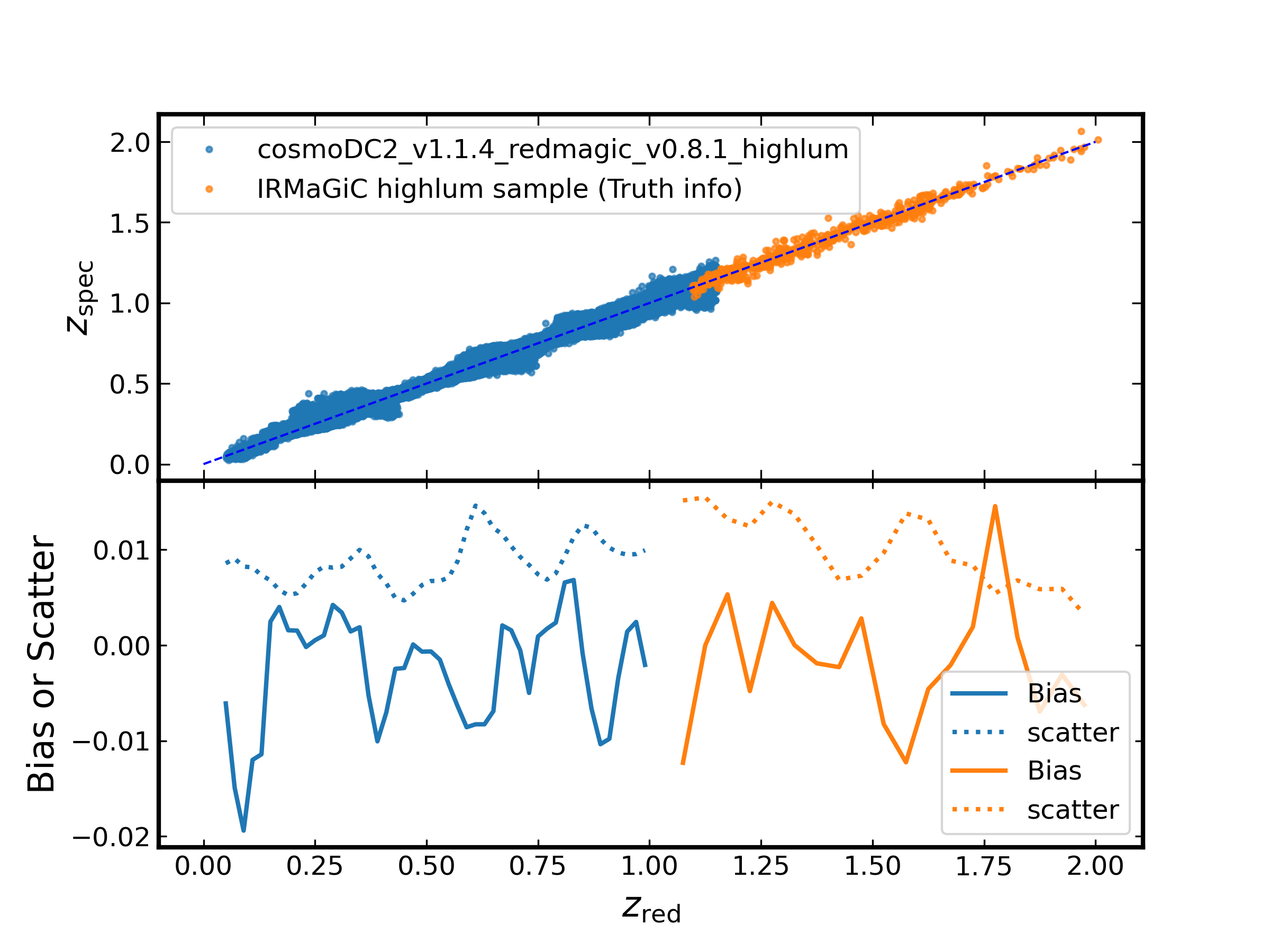}
    \caption{Photometric redshift performance comparison of the cosmoDC2 RedMaGiC samples and LRG samples derived based on truth catalog in this study for the highdens and higlum samples relatively. Top panels shows $z_\mathrm{red}$ versus $z_\mathrm{spec}$. The bottom panels shows bias (solid line) the NMAD (dotted line).}
    \label{fig:redmagic-truth-cat-compare}
\end{figure*}

In Section~\ref{sec: rs-caliberation}, we reported unusually high intrinsic scatter across the spline nodes for the $z - Y$ color. Moreover, in Section~\ref{sec: compare with DC2 redmagic}, the LRG catalog generated in this study exhibited greater scatter than that of the low-redshift DC2 RedMaGiC catalog. This discrepancy might be attributed to the photometric measurements in the Roman galaxy sample. To further explore this issue, we utilized the galaxy truth photometry (specificially, $i$ and $z$ magnitude for LSST and $Y,\ J,\ H,\ F$ magnitude for Roman) from input truth catalogs to calibrate a red-sequence model based on ground-truth information and compare the calibration results and RedMaGiC redshift performance. For the Roman data, the truth catalog provided by \citet{Troxel23}, for the LSST data, we cross-match the galaxy id from the Roman truth catalog with the cosmoDC2 catalog, \verb|cosmoDC2_v1.1.4_image|. The original RedMaGiC algorithm was applied to the cosmoDC2 sample to generate LRG samples, \verb|cosmoDC2_v1.1.4_redmagic_v0.8.1_highdens| and \verb|cosmoDC2_v1.1.4_redmagic_v0.8.1_highlum|, using LSST-only truth photometry (g,r,i,z,y) in the redshift range of $0 < z < 1$. We compare our results with the cosmoDC2 LRG samples in  Figure~\ref{fig:redmagic-truth-cat-compare} and the bottom panel displays the scatter as a function of $z_{\text{red}}$. For comparison of both the highdens and highlum samples, the scatter in the high-redshift LRG samples from this study is comparable to that in the low-redshift cosmoDC2 LRG samples. This suggests that the observed increased scatter in high-redshift LRG samples is not due to deficiencies in the RedMaGiC algorithm but rather calibration issues with the photometry of the Roman simulated galaxy sample.

\end{document}